\documentclass{emulateapj}
\newcommand{\HI}{\ion{H}{1}~}

\newcommand{\kms}{km~s$^{-1}$ }
\newcommand{\Lya}{Lyman~$\alpha$~}

\usepackage{lscape} 
\usepackage{color} 
\slugcomment{Accepted for publication in ApJ: Sep 20, 2016}

\shorttitle{CGM in Blue and Red Galaxies}
\shortauthors{Borthakur et al.}

\begin{document}

\title{The Properties of the Circumgalactic Medium in Red and Blue Galaxies: Results from the COS-GASS$+$COS-Halos Surveys }

\author{Sanchayeeta Borthakur}
\affil{Department of Physics \& Astronomy, Johns Hopkins University, Baltimore, MD, 21218, USA; The Astronomy Department, University of California, Berkeley, CA, USA}
\email{sanch@pha.jhu.edu}

\author{Timothy Heckman}
\affil{Department of Physics \& Astronomy, Johns Hopkins University, Baltimore, MD, 21218, USA}

\author{Jason Tumlinson}
\affil{Space Telescope Science Institute, Baltimore, MD, USA}

\author{Rongmon Bordoloi}
\affil{MIT-Kavli Center for Astrophysics and Space Research, Cambridge, MA, USA} 

\author{Guinevere Kauffmann}
\affil{Max-Planck Institut f$\rm\ddot{u}$r Astrophysik, D-85741 Garching, Germany }

\author{Barbara Catinella}
\affil{International Centre for Radio Astronomy Research, The University of Western Australia, 35 Stirling Highway, Crawley WA 6009, Australia}

\author{David Schiminovich}
\affil{Department of Astronomy, Columbia University}

\author{Romeel Dav\'e}
\affil{University of the Western Cape, Bellville, Cape Town 7535, \\
South Africa, South African Astronomical Observatories, Observatory, Cape Town 7925, \\
South Africa, African Institute for Mathematical Sciences, Muizenberg, Cape Town 7945, South Africa}

\author{Sean M. Moran}
\affil{Harvard-Smithsonian Center for Astrophysics, 60 Garden Street, Cambridge, MA 02138}

\author{Amelie Saintonge}
\affil{Department of Physics and Astronomy, University College London, Gower Place, London WC1E 6BT, UK}


\begin{abstract}

We use the combined data from the COS-GASS and COS-Halos surveys to characterize the Circum-Galactic Medium (CGM) surrounding typical low-redshift galaxies in the mass range $\rm~M_*\sim~10^{9.5-11.5}~M_{\odot} $, and over a range of impact parameters extending to just beyond the halo virial radius ($\rm~R_{vir}$). We find the radial scale length of the distributions of the equivalent widths of the \Lya and \ion{Si}{3} absorbers to be 0.9 and 0.4 $\rm~R_{vir}$, respectively. 
The radial distribution of equivalent widths is relatively uniform for the blue galaxies, but highly patchy (low covering fraction) for the red galaxies. We also find that the \Lya and  \ion{Si}{3} equivalent widths show significant positive correlations with the specific star-formation rate (sSFR) of the galaxy. We find a surprising lack of correlations between the halo mass (virial velocity) and either the velocity dispersions or velocity offsets of the \Lya lines. 
The ratio of the velocity offset to the velocity dispersion for the \Lya absorbers has a mean value of $\sim$ 4, suggesting that a given the line-of-sight is intersecting a dynamically coherent structure in the CGM rather than a sea of orbiting clouds. The kinematic properties of the CGM are similar in the blue and red galaxies, although we find that a significantly larger fraction of the blue galaxies have large \Lya velocity offsets ($>$~200~\kms). 
We show that - if the CGM clouds represent future fuel for star-formation - our new results could imply a large drop in the specific star-formation rate across the galaxy mass-range we probe.

\end{abstract}

\section{INTRODUCTION\label{intro}}

Galaxy growth is fundamentally connected to the cycle of accretion and ejection of matter into and out of galaxies. In the simplest picture, galaxies acquire gas that reaches the central regions via the circum-galactic medium (CGM). There it condenses into neutral and then molecular gas, some of which is then converted into stars. Young stars in turn drive strong winds, outflows, and radiation that deposit mass, metals, energy, and momentum to the CGM, thus significantly influencing its properties \citep[see review by][and references therein]{somerville15,fielding16}. These linked processes are commonly termed the baryon cycle. The CGM then lies at the heart of this cycle, as it is the interface between the stellar body of the galaxy and the intergalactic medium. It is the primary spatial pathway for the baryon cycle into and out of galaxies \citep[][and references therein]{ford16, tumlinson13, borthakur15, nielsen15, shen14, mitra15, brook14}. 

The CGM is also a reservoir of low-density gas that may have as much mass as the stellar component of the galaxy \citep{werk13,werk14, tumlinson13, peeples14, richter16}. It extends out from the stellar disk out to the virial radius of the galaxy \citep{chen01a, stocke13, borthakur13}. However, due to its low surface-brightness, we have not yet been able to directly image this vast baryonic reservoir. On the other hand, absorption-line spectroscopy provides an avenue to probe the physical conditions in this low-density gaseous medium. Rest-frame ultra-violet (UV) spectroscopy enables us to use various absorption-line transitions, including hydrogen and metal-line species spanning a broad range of ionization states. 

Mapping the CGM with the help of a large sample of sightlines probing a range of impact parameters is crucial for understanding its properties and its variations as a function of radius. The radial dependence in the properties of the neutral hydrogen in the CGM has been known for decades, based on observations of the \Lya absorption-line \citep[][and references therein]{lanzetta95, chen98, tripp98, chen01b, bowen02, prochaska11, stocke13, tumlinson13, liang14, borthakur15}. However, only recently, with the installation of Cosmic Origins Spectrograph (COS) aboard the Hubble Space Telescope (HST), has it become feasible to undertake detailed probes of the CGM properties as a function of other global properties of the central galaxy.  

One of the consequences of the accretion of gas passing through the CGM is that this provides the raw material to sustain the growth of the galaxy via star formation \citep[e.g.][]{bouche13}. Not all galaxies produce stars at the same rate \citep{brinchmann04, salim07, noeske07, daddi07, rodighiero11,  speagle14,snyder15}. In particular, galaxies show two distinct populations in terms of their star-formation rate (SFR). While most low mass galaxies form stars at significant rates, most high mass galaxies produce stars at negligible levels. This was termed as the galaxy color bimodality defined in terms of ``blue" (star-forming) galaxies and ``red" (quiescent) galaxies \citep[e.g.][]{kauffmann03, blanton03, baldry04, brinch04, tully82}.

About a decade back, cosmological hydrodynamical simulations revealed two distinct ways that galaxies accrete gas into their dark matter halo as a function of halo mass. The predominant mode of gas accretion for low mass galaxies is believed to be the ``cold" mode \citep{keres05, keres09, dekel09}, where gas falls into galaxies as streams or lumps at temperatures much less than that of the virial temperature. For the higher mass halos, the accretion process is expected to be in the ``hot" mode \citep{white91, fukugita06}, in which the incoming gas shock heats to the virial temperature. This broadly can explain why high-mass galaxies have little to no cold gas reservoirs to fuel star-formation \citep[see work on condensation in hydrodynamical simulations by ][]{kaufmann06, kaufmann09, sommerlarsen06}. Galaxies also recycle gas from previous generations of star-formation that is stored in their CGM \citep{ford13,fraternali15}.
 However, the process of how gas gets into the disk from the CGM is fairly complex. 
Nonlinear perturbations in the filamentary flows may help the cool accreting gas condense and add cold gas to the disk \citep{keres_hernquist09, joung12}. These condensing clouds may contain as much as  25\%-75\% of the cold gas in the CGM \citep{fernandez12}.

\begin{figure*}
\includegraphics[trim = 20mm 105mm 20mm 0mm, clip,scale=0.535,angle=-0]{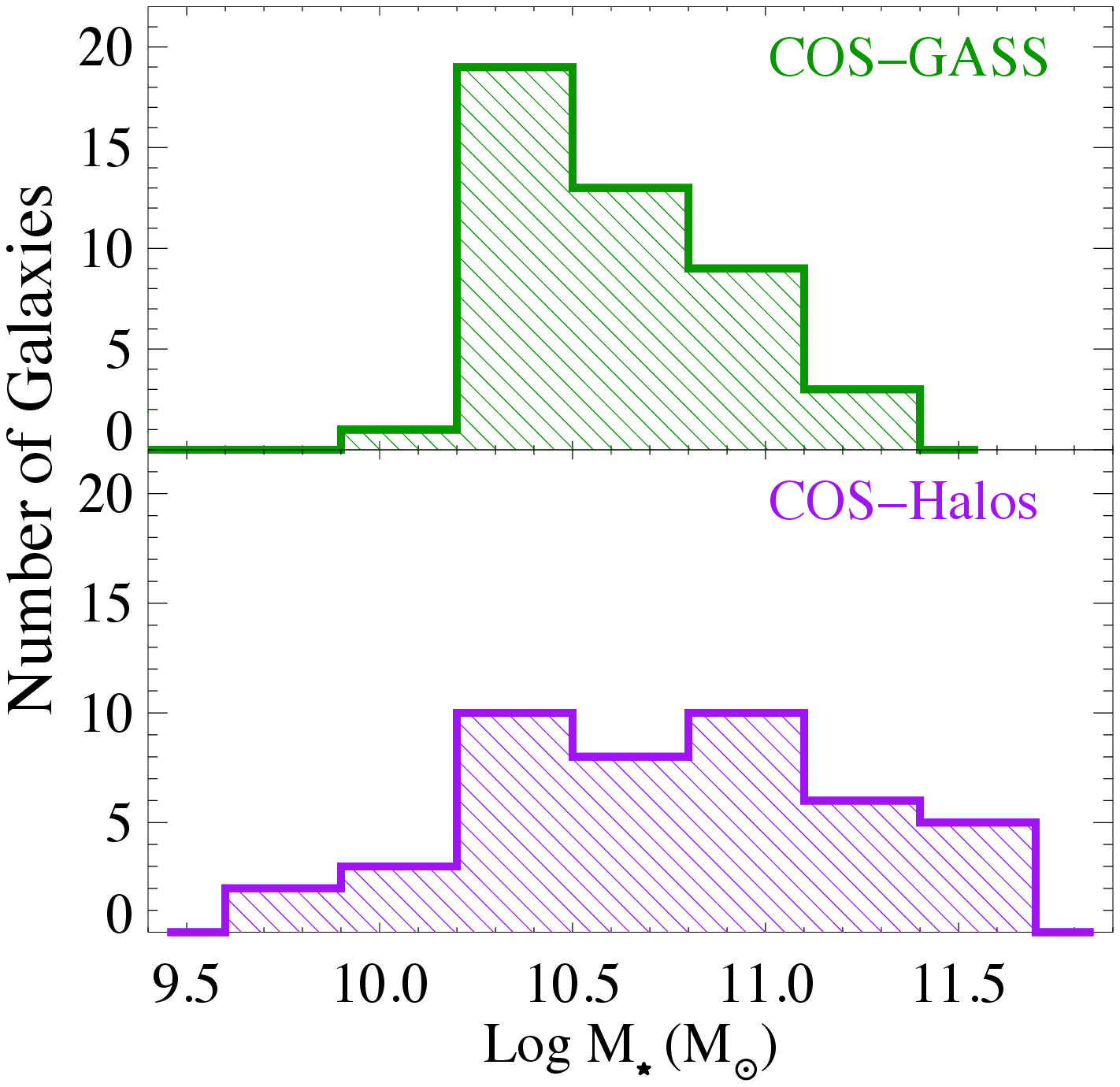}   
\includegraphics[trim = 20mm 105mm 20mm 0mm, clip,scale=0.535,angle=-0]{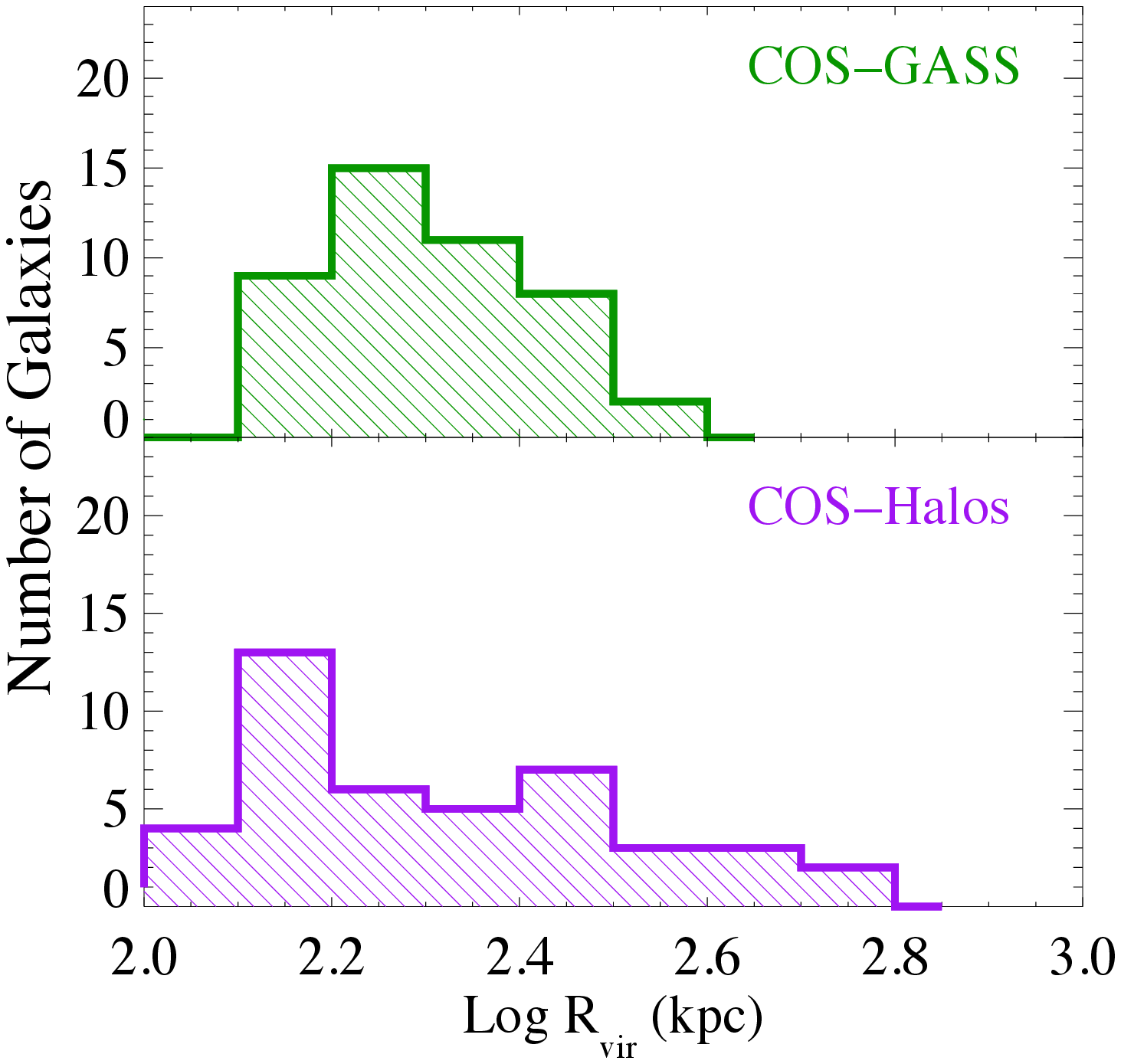}    
\caption{Distribution of galaxy properties for the COS-GASS and COS-Halos samples. The left panel shows the stellar mass distribution and the right panel shows the virial radius distribution. The $\rm R_{vir}$ for both the samples were estimated using the prescription described by \citet{kravtsov14, mandelbaum16,liang14} as described in section 2.1.}
 \label{fig-sample_m_Rvir} 
\end{figure*}

 \begin{figure*}
 \hspace{-0.6cm}
\includegraphics[trim = 0mm 0mm 0mm 0mm, clip,scale=0.535,angle=-0]{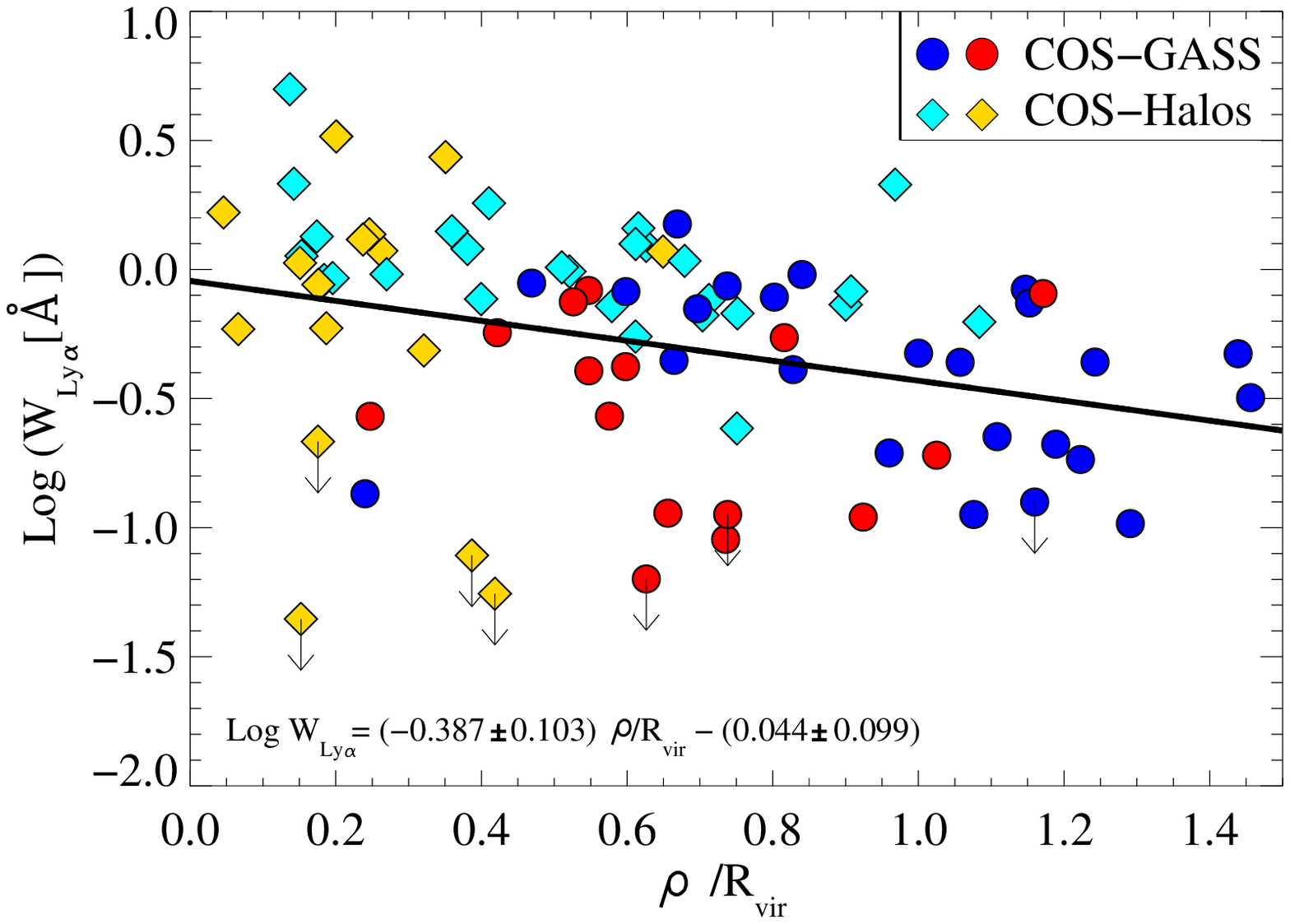} 
\includegraphics[trim = 0mm 0mm 0mm 0mm, clip,scale=0.535,angle=-0]{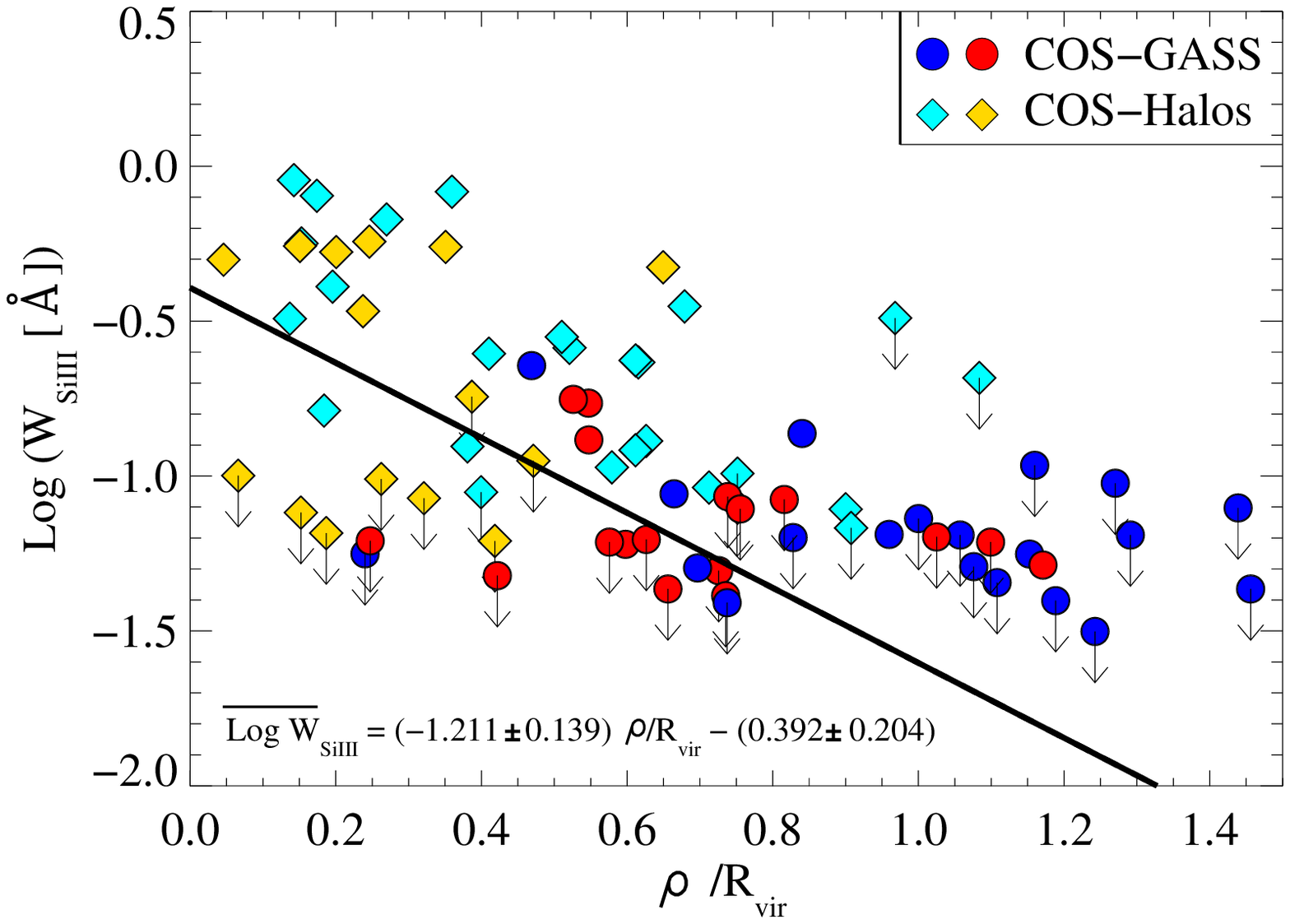} 
\caption{Variation of \Lya and Si~III  equivalent width with normalized impact parameter (i.e. $\rm \rho/R_{vir}$) for a combined COS-GASS and COS-Halos sample. The colors blue and cyan indicate``blue" galaxies and red and yellow denote ``red" galaxies. The black thick line denotes the fits to the data using the Buckley-James method. The calculations were performed using the survival analysis software ASURV that takes into account the censored data. The parameters describing the best-fit lines are printed at the bottom left corner. Since the fits presented here take into account the censored data, the parameters of the best-fit in the left panel are slightly different from those published by \citet{borthakur15}.}
 \label{fig-W_rho_Rvir} 
\end{figure*}

In addition to accretion, star-formation driven feedback may change the nature and properties of the gas in the CGM \citep{kauffmann16, liang16, nelson15, nelson16, marasco15}. Massive young stars inject energy and/or momentum into outflows \citep{veilleux05, heckman11, borthakur14, heckman15, heckman16} that may travel into the CGM, enriching it with metals, shock-heating the cooler CGM clouds, and possibly even expelling/unbinding the CGM \citep{borthakur13}. Therefore, if feedback provided by massive stars plays a role in the observed bimodality, then we should see a change in the structure, ionization state, and/or kinematics of the CGM as a function of SFR.

To that end, we have selected a subsample of galaxies from the GALEX Arecibo SDSS Survey \citep[GASS;][]{catinella10,catinella12,catinella13} that have background UV-bright quasi-stellar objects (QSOs) located within a projected distance of 250~kpc in the rest-frame of the galaxy. This yielded the COS-GASS sample \citep{borthakur15} whose members were observed with COS using the G130M grating. This provided a spectral R$ =$ 20,000$-$24,000 (FWHM $\sim$ 12 to 15~\kms). We have multi-band data for these galaxies from the parent GASS survey:  21~cm \HI spectroscopic data obtained with the Arecibo telescope, optical images and spectroscopy from the Sloan Digital Sky Survey (SDSS), UV imaging with the Galaxy Evolution Explorer (GALEX), molecular gas data from IRAM \citep[COLD GASS;][]{saintonge11}, and long-slit optical spectroscopy \citep{moran12} for a portion of the sample. Therefore, we have the stellar mass, SFR, gas-phase metallicity, stellar morphology, and atomic and molecular gas masses for all the 45 galaxies from the COS-GASS sample.

Here we present our study utilizing the combined COS-GASS \citep{borthakur15} and COS-Halos \citep{tumlinson13} samples. Detailed descriptions of our sample, the COS observations and data reduction are presented in Section~2.  The results are presented in Section~ 3 and their implications are discussed in Section~4. Finally, we summarize our findings in Section~5. The cosmological parameters used in this study are $H_0 =70~{\rm km~s}^{-1}~{\rm Mpc}^{-1}$ (in between the two recent measurements of $\rm 73.24 \pm 1.74 ~ km~s^{-1}~Mpc^{-1}$ \citep{riess16} and $\rm 67.6_{-0.6}^{+0.7}~{\rm km~s}^{-1}~{\rm Mpc}^{-1}$ \citep{grieb16}) , $\Omega_m = 0.3$, and $\Omega_{\Lambda} = 0.7$. We note that varying the Hubble constant value from $\rm 65~ to~ 75 ~ km~s^{-1}~Mpc^{-1}$ does not affect the conclusions in the paper.


\begin{figure}
\hspace{-0.2cm}
\includegraphics[trim = 0mm 0mm 0mm 0mm, clip,scale=0.5, angle=-0]{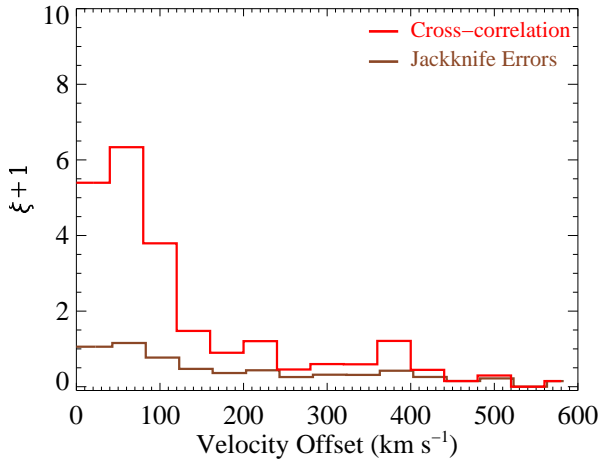} 
\caption{ The cross-correlation function between the galaxy systemic velocities and velocity centroids of \Lya absorbers. The cross-correlation function was calculated using the same data analysis criterion as the observations/measurements. For example, the absorbers were randomly distributed within the allowed velocity range of $\pm$600~\kms. Caution must be applied when comparing these results to those from blind surveys or surveys with different intrinsic resolutions for the spectrograph.    }
 \label{fig-cc_lya} 
\end{figure}

\section{OBSERVATIONS  \label{sec:observations}}

\subsection{Sample \label{sec:sample}}

We combined the COS-GASS sample (45 galaxies) with the COS-Halos sample (44 galaxies) to get our full sample. The COS-Halos sightlines cover the inner CGM (thoroughly up to $\sim$0.8$\rm R_{vir}$) whereas the COS-GASS sightlines extend the observations to the outer CGM ($\sim$ 0.05-1.5$\rm R_{vir}$). The resulting combined sample contains a total of 89 sightlines probing the CGM from 17-231~kpc in the rest frame of the target galaxies. 

The two programs probe similar stellar mass ranges. The COS-Halos program probes galaxies in the range $ \rm 10^{9.6-11.5}~M_{\odot}$ at $\rm 0.1<z< 0.2$ whereas the COS-GASS program probes galaxies in the range ($\rm 10^{10.1-11.1} ~M_{\odot}$) at slightly lower redshifts of 0.02$<z<$0.05. A comparison of the stellar masses and virial radii (based on the prescription by \citet{kravtsov14} and \citet{liang14}) for both samples are provided in Figure~\ref{fig-sample_m_Rvir}. Recent gravitational-lensing-based results by \citet{mandelbaum16} show that blue galaxies of fixed stellar mass are found in lower mass halos than red galaxies of the same stellar mass. Based on their results, we add (subtract) 0.15~dex to the halo masses for red (blue) galaxies with a given stellar mass. The dark matter halo masses of the combined sample range from 11.1 to 13.2~$\rm M_{\odot}$ dex. We note that the redshift difference between the two samples is $\sim$ 0.1. However, the variation in CGM properties during this time is expected to be minimal \citep{chen12}.
 Also, the COS-Halos sample was selected to be all centrals \citep[with a couple of non-centrals][section 2.5]{tumlinson13}. This is not one of the criteria for COS-GASS, although the mass range of the galaxies ensured that most of the COS-GASS galaxies (34/45) are centrals \citep[based on the group catalog by][]{yang05,yang07}. So the combined sample is 85\% centrals. We have retained the satellites in our analysis, but have verified that they do not affect any of our conclusions.

We identify galaxies with specific star formation rates (sSFR= SFR/M$_{\star}) >\rm 10^{-11}~yr^{-1}$ as blue (star-forming) galaxies and those below this limit as red (quiescent) galaxies. sSFR values of $<\rm 10^{-12}~yr^{-1}$ can be considered as upper limit. A detailed description of our galaxy color assignment can be found in \citet{borthakur15}.

\begin{figure}
\hspace{-0.2cm}
\includegraphics[trim = 0mm 0mm 0mm 0mm, clip,scale=0.5, angle=-0]{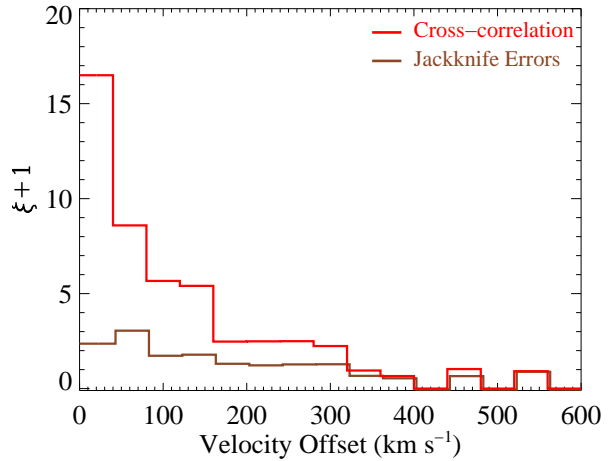} 
\caption{ The cross-correlation function between velocity centroids of \Lya  and Si~III absorbers. The calculations for the cross-correlation function preserved the distribution of the \Lya absorbers with respect to the galaxy systemic velocity as discussed in Figure~\ref{fig-cc_lya}.   }
 \label{fig-cc_Si3}
  \end{figure}

For more information on the properties of the target galaxies, including their redshifts, stellar masses, SFRs\footnote{
The SFR for the GASS sample were derived by combining GALEX $FUV$ and $NUV$ and SDSS $u,g,r,i,z$ photometry and SDSS spectral-line indices.}, sSFR, galaxy colors, and impact parameter of the sightlines, we refer the reader to Table~1 presented in this paper (for the COS-GASS sample) and Table~2 from the published work by \citet{tumlinson13}.


 \subsection{Observations and Data Reduction \label{sec:observations_datareduction}}

 The data presented in this paper were obtained under the COS-GASS survey (program=12603; P.I. Heckman) observed with the COS aboard the HST using the high resolution grating G130M (R= 20,000-24,000 ; FWHM=12-15\kms).
The wavelength coverage of the spectrograph is 1140-1470~$\rm \AA$. 
The galaxies being at lower redshift  (maximum redshift of 0.05) allows us to probe a wide variety of far-UV line transitions such as \Lya ($\rm \lambda 1216 $), \ion{Si}{2}~($\rm\lambda 1190, 1193, \& 1260$), \ion{Si}{3}~($\rm \lambda 1206$), \ion{Si}{4}~($\rm \lambda\lambda1393, 1402$), \ion{C}{2} ($\rm \lambda 1334$), and \ion{O}{1}~($\rm \lambda 1302$). 
 
Absorption features that have equivalent widths larger than 3 times the noise in the spectra were picked out and then identified both in terms of the transition and redshift. This allowed us to detect any contamination to the absorption associated with the target galaxies. We searched in a velocity window of $\pm \rm 600$~\kms from the systemic velocity (using optimal redshift from SDSS that is tracing the stars and ionized gas in the central region of the galaxies) for associated absorbers. The absorbers were measured and Voigt profile fits were performed. More information on the data reduction can be found in the previous publication of COS-GASS \citep{borthakur15}. This procedure is exactly same as followed by \citet{tumlinson13, werk13} for the COS-Halos program.


\section{Results \label{sec:Gass_halos}}

The COS-GASS survey covered a wavelength range of $\approx$1150-1450$\rm \AA$ for most galaxies. This includes the prominent transitions like \HI \Lya $\lambda 1216$, \ion{O}{1} $\lambda 1302$, \ion{C}{2} $\lambda 1334$, \ion{Si}{2} $\lambda 1260, 1193, 1190$, 
\ion{Si}{3}$\lambda 1206$, \ion{Si}{4}$\lambda \lambda 1393,1402$,  and \ion{N}{5}$\lambda 1239$. \HI \Lya and \ion{Si}{3} $\lambda 1206$ are the strongest lines detected in our sample. In this paper, we will primarily focus on these two most sensitive probes. A paper presenting all the other metal-lines detected in COS-GASS survey is in preparation.
Table~2 presents the measurements for \Lya and \ion{Si}{3} for each of the sightlines from the COS-GASS sample. In cases where we do not detect any absorption features, we quote a 3$\sigma$ equivalent width as the upper limit.
The detection limits for the COS-GASS sample are typically $\sim \rm 50~m\AA$, which corresponds to Log~N(\ion{H}{1})= 12.96, Log~N(\ion{Si}{2})= 12.55, Log~N(\ion{Si}{3})= 12.37, Log~N(\ion{Si}{4})= 13.05, and Log~N(\ion{C}{2})= 13.39, respectively. 

The measurements for COS-Halos sightlines can be found in the published work by \citet{werk13}.

\begin{figure}
\hspace{-0.5cm}
\includegraphics[trim = 15mm 5mm 5mm 0mm, clip,scale=0.6, angle=-0]{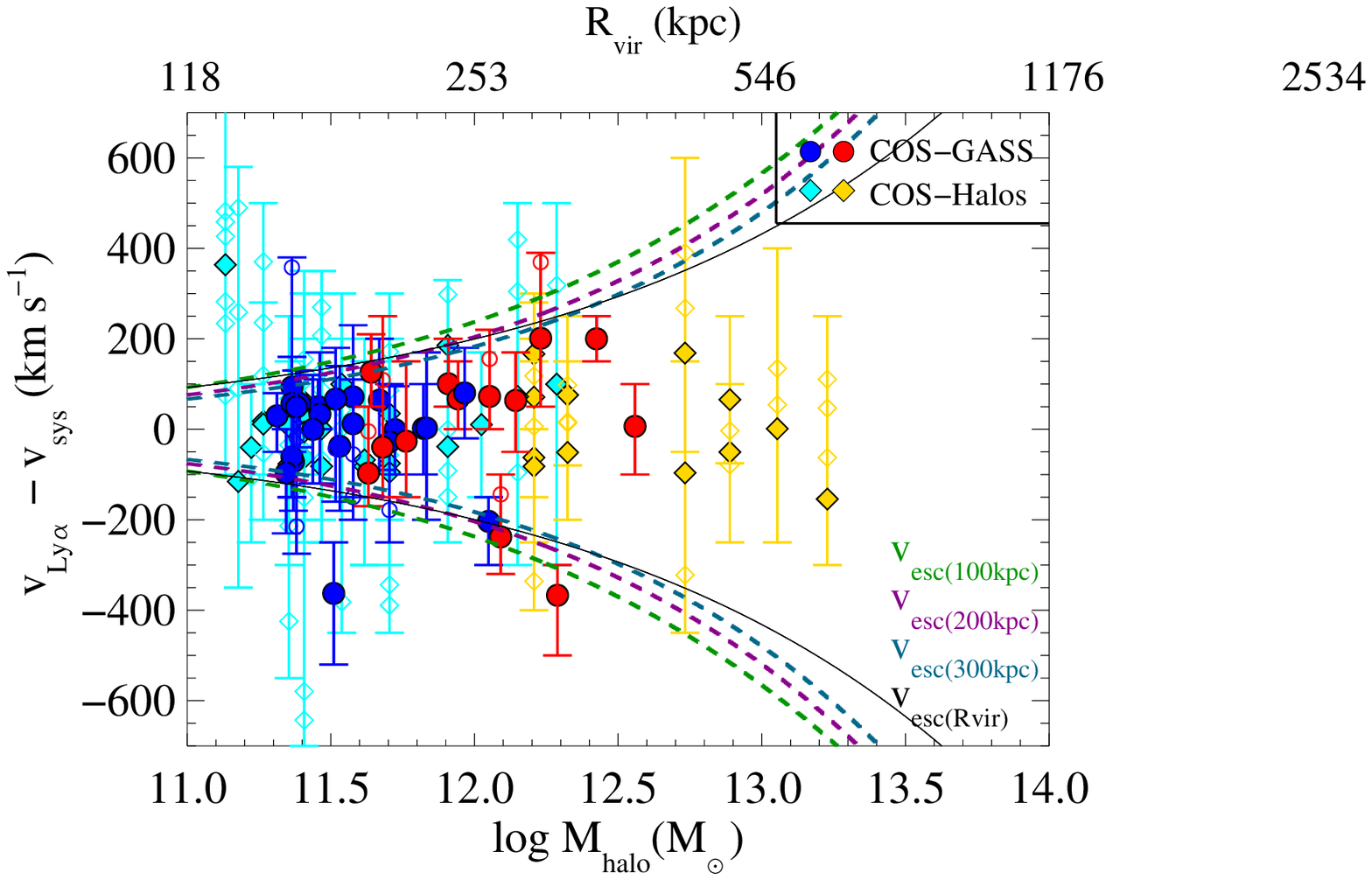} 
\caption{ Velocity distribution of \Lya transitions with respective to the systemic velocity of the host galaxies for the COS-GASS and COS-Halos samples. The color of the symbols show the color of the galaxy (blue and cyan for ``blue" galaxies and red and yellow for ``red" galaxies) and the vertical colored bar shows the extent of absorption. The centroids of the strongest component are shown as the filled symbols and the weaker components are shown as open symbols. The escape velocity required for the gas clouds to escape the halos at impact parameters of 100, 200, and 300~kpc are shown as colored dashed curves and that at the virial radius as solid black curve. The velocity distribution of metal species is very similar to this plot, although the constraints are weaker due to multiple non-detections.  }
 \label{fig-vel_mass_esc}
\end{figure}

\begin{figure}
\includegraphics[trim = 15mm 0mm 5mm 0mm, clip,scale=0.55, angle=-0]{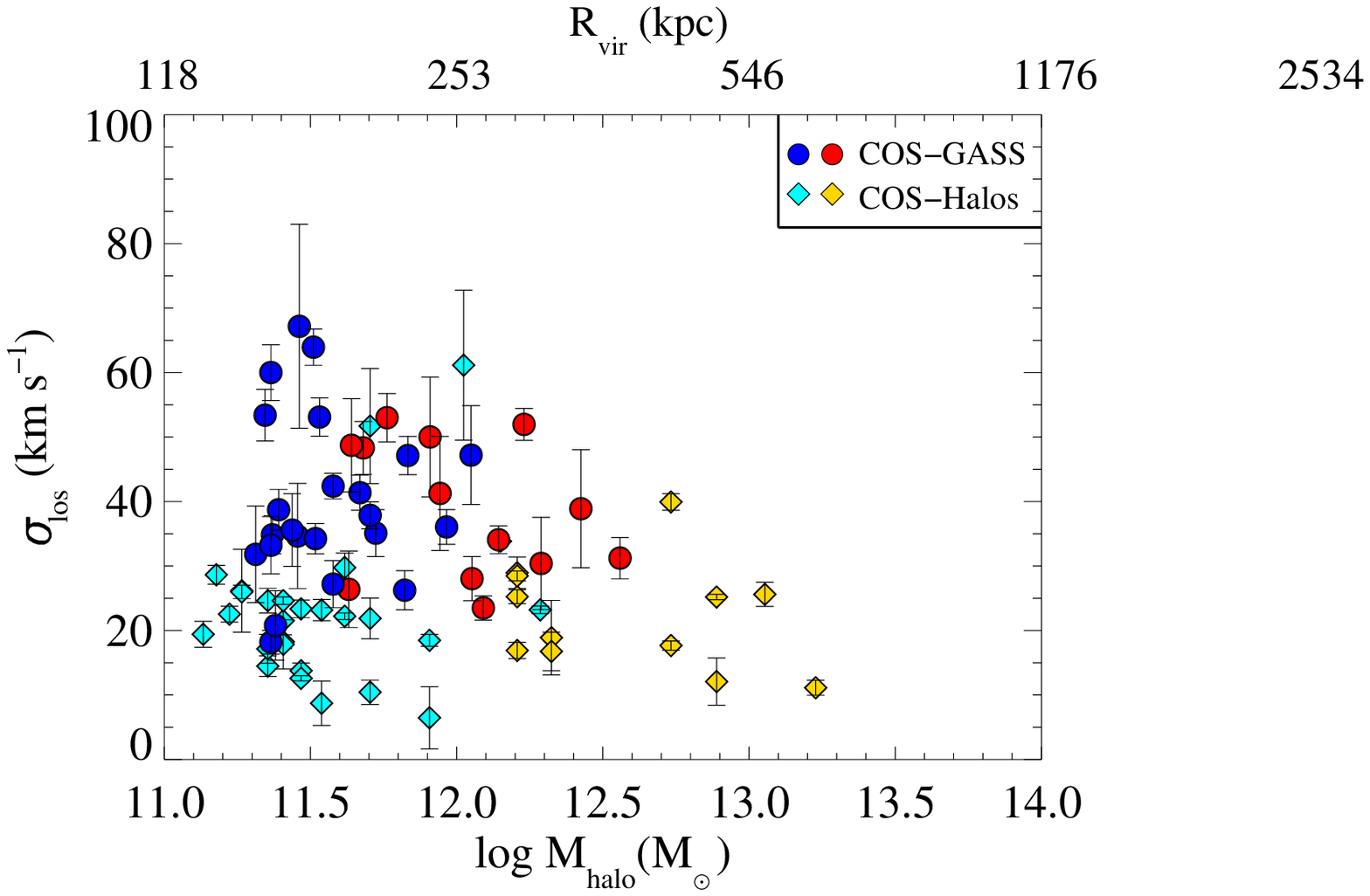} 
\caption{The line-of-sight velocity dispersion for the strongest component of the \Lya absorption line plotted as a function of the halo mass. There is no significant correlation between these parameters.}
 \label{fig-sigma_mhalo} 
\end{figure}

\begin{figure}
\includegraphics[trim = 15mm 0mm 5mm 0mm, clip,scale=0.55, angle=-0]{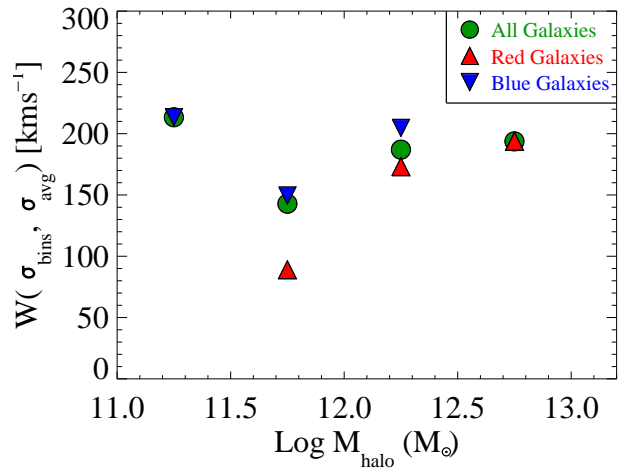} 
\caption{ Velocity parameter, W, as a function of halo mass. The parameter W defined as $\rm W^2= \sigma_{bin}^2 + \sigma_{avg}^2 $ where $\rm \sigma_{bin}$ is the  dispersion of the centroids within each bin  and the $\rm \sigma_{avg}$ is the average velocity dispersion (width) among the absorption features within the bins. The full sample is shown in green and the red and blue galaxies are shown as red and blue triangles.  The value of W does not change significantly between halo masses of $10^{11-13}$ M$_{\odot}$ for the entire sample.}
 \label{fig-vel_mass} 
\end{figure}

\begin{figure*}
\includegraphics[trim = 15mm 5mm 0mm 15mm, clip,scale=0.6, angle=-0]{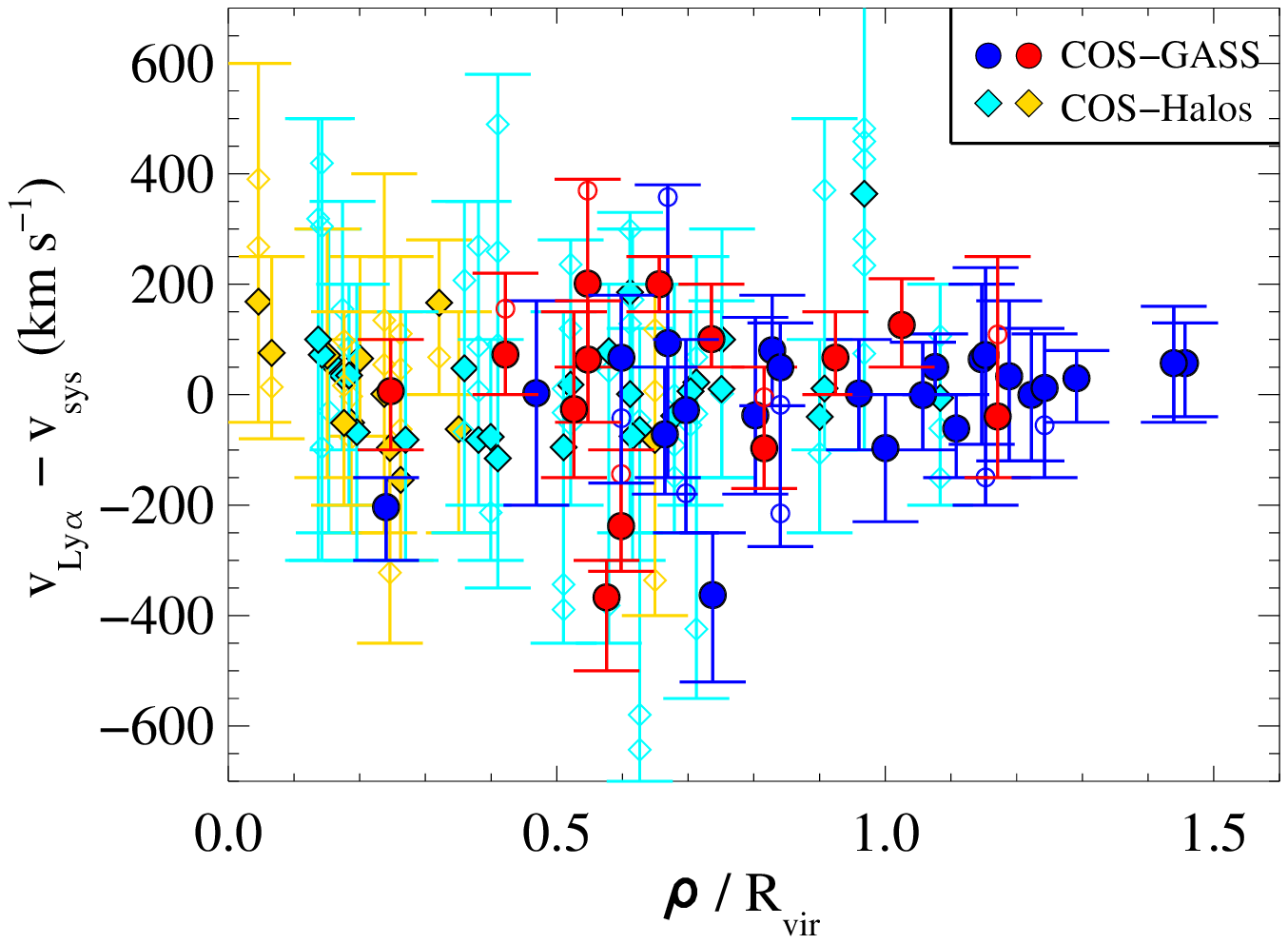}   
\includegraphics[trim = 15mm 0mm 0mm 0mm, clip,scale=0.53, angle=-0]{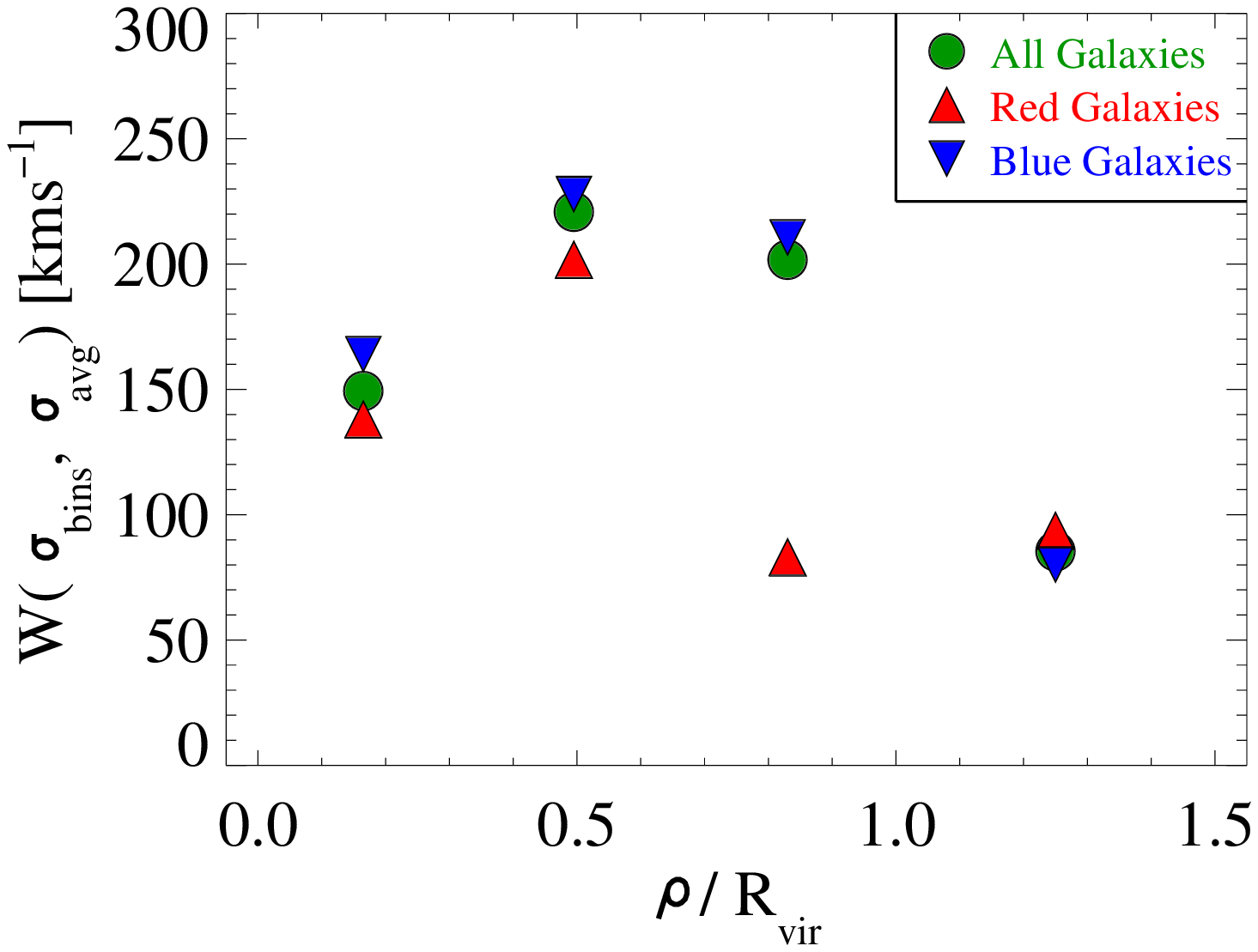}   
\caption{Velocity distribution of \Lya transitions with respective to the systemic velocity of the host galaxies as a function of normalized impact parameter ($\rm \rho/R_{vir}$) for the COS-GASS  and COS-Halos samples. The circles represent the COS-GASS sample and the diamonds represent the COS-Halos sample. The color of the symbols indicate the color of the galaxy i.e. blue and cyan for ``blue" galaxies and red and yellow for ``red" galaxies.}
 \label{fig-vel_rho_Rvir} 
\end{figure*}

\subsection{Overall Detection Rates \label{sec:dect_rates}}

The \Lya absorption-line, produced by neutral Hydrogen, is the strongest transition found in the combined data. \Lya absorption was detected in 75/82 ($\rm  91^{+9}_{-15}\%$) sightlines where measurements could be made. This detection rate is consistent with those found by \citet{prochaska11,stocke13, liang14}.
Often the absorption features are saturated and hence we use the equivalent width for our analysis, as we are not able to accurately determine the column densities of these absorbers. 
The \Lya is primarily tracing  $\rm \approx 10^{4-5.5}~K$ (based on the Doppler widths\footnote{This does not rule out a the presence of a substantial ``hot" medium at temperatures of $\approx \rm T > 10^6~K$ that may be traced by species like \ion{O}{7}}).

The \ion{Si}{3} $\lambda$1206.5 absorption-line is the strongest feature tracing metals in the COS-GASS survey, thus making it the most sensitive tracer of the warm CGM \citep[also see][]{collins09,  shull09, lehner12, lehner15, richter16}. Out of 37 sightlines for the COS-GASS sample, where data were uncontaminated and measurements could be made, we detected \ion{Si}{3} in 11 of them. Thus the detection rate of \ion{Si}{3} in the outer halo is 30$\pm$10\%. This is smaller than results for the inner CGM, as found by \citet[COS-Halos survey;][]{werk13}, \citet{liang14} and \citet{richter16}\footnote{It is worth noting that the study by \citet{richter16} is a statistical study of the \ion{Si}{3} towards 303 QSO sightlines that may be associated with galaxies (instead of a targeted study of the CGM)}.
Since our study primarily focuses on the outer CGM ( 0.5$\rm R_{vir} < \rho < 1. 5 R_{vir}$ with the exception of two inner sightlines), a more relevant comparison would be the covering fraction of \ion{Si}{3} of $14^{+11}_{-5}$\% for sightlines with 0.54~$\rm R_{vir} < \rho < 1. 02~R_{vir}$ by \citet{liang14}. Both the numbers are broadly consistent given that the error ranges in the estimate do have an overlap. 
Interesting, the two inner sightlines in the COS-GASS sample, which one might naively think are contributing the excess in our detection rate are devoid of any \ion{Si}{3} absorption. However, in the broad context of metals in the outer CGM, our study find consistently lower metal covering fraction thus suggesting that metals are rarer in the outer CGM similar to conclusions of \citet[][]{liang14} and \citet[][for \ion{C}{4} from the COS-Dwarfs survey]{bordoloi14}. The combined sample has a detection rate of \ion{Si}{3} of $\rm 49\pm10$\%.

The detection rate of \ion{C}{2}, \ion{Si}{2} and \ion{Si}{4} in the COS-GASS sample is 20$\pm$8\%, 7$\pm$5\%, and 9$\pm$5\% respectively. These are much smaller than that of the detection rate of \ion{Si}{3}, although we note that in some sightlines the \ion{Si}{2} $\rm \lambda~1260~\AA$was corrupted by the geocoronal \ion{O}{1} emission feature and hence suffer from small number statistics.  
Our sensitivity is higher for \ion{Si}{3} as compared to similar columns of \ion{Si}{2} and \ion{Si}{4}. Therefore, a fair comparison is to compare the detection rate at the same column density. For example, we detected 6/31 (16$\pm$7\%) \ion{Si}{3} absorbers with equivalent widths above the  0.077$\rm \AA$ corresponding to a column density of Log~N(\ion{Si}{3})=12.55. At this same column density sensitivity, the detection rate of \ion{Si}{2} is less than half that of \ion{Si}{3}. However, the same argument cannot be applied to \ion{Si}{4} as our sensitivity to \ion{Si}{4} is about an-order-of-magnitude lower, although none of the \ion{Si}{3} absorbers have associated \ion{Si}{4} absorption.

Most of the \ion{Si}{3} absorbers are tracing warm intermediate ionization circumgalactic gas. 
We find the observed line ratios of \ion{Si}{2}, \ion{Si}{3}, \ion{Si}{4}, and \Lya from the COS-GASS sample to be consistent with photoionization of the CGM by the cosmic ultraviolet background. We expect the CGM in the outer halo as traced by \ion{Si}{3} to have an ionization parameter, U, of -2.8$< \rm log~U < $ -1.7, although the exact upper bound is hard to set given the saturation of \Lya and non-detection of \ion{Si}{4}. Similar ionization parameters were also estimated by \citet{shull09} for \ion{Si}{3} associated with the high- and intermediate-velocity clouds (HVC, IVC) in the Milky Way halo \footnote{The conditions may not exactly be similar between the HVC/IVC and the COS-GASS absorbers since the HVC/IVC are within 50~kpc of the Milky Way disk \citep{lehner12} and not the outer CGM \citep[see][for more on vantage point correction]{richter12,herenz13}}. 
This value of ionization parameter is lower that required to produce a substantial amounts of \ion{O}{6} and \ion{C}{4} absorbers for the observed \Lya column densities of $\rm 10^{14-15}~atoms~cm^{-2}$ that are seen in our sample. 
And therefore, it is likely that most of these highly ionized absorbers are different from those detected in various other CGM and QSO-absorption studies \citep[][and references therein]{tumlinson11b, chen01a, wakker09,  borthakur13, bordoloi14}.  
However, as discussed in detail by \citet[][and references therein]{werk14, fox13,  meiring13, lehner13, tripp11}, \ion{O}{6} may represent a different phase of gas that differs from the ones traced by lower ionization transitions. 
Since the COS-GASS data do not cover the \ion{O}{6} line, we refrain from further discussion of \ion{O}{6}. Instead, we focus on \Lya and \ion{Si}{3} in the remainder of the paper.

\begin{figure*}
\includegraphics[trim = 0mm 0mm 35mm 0mm, clip,scale=0.65,angle=-0]{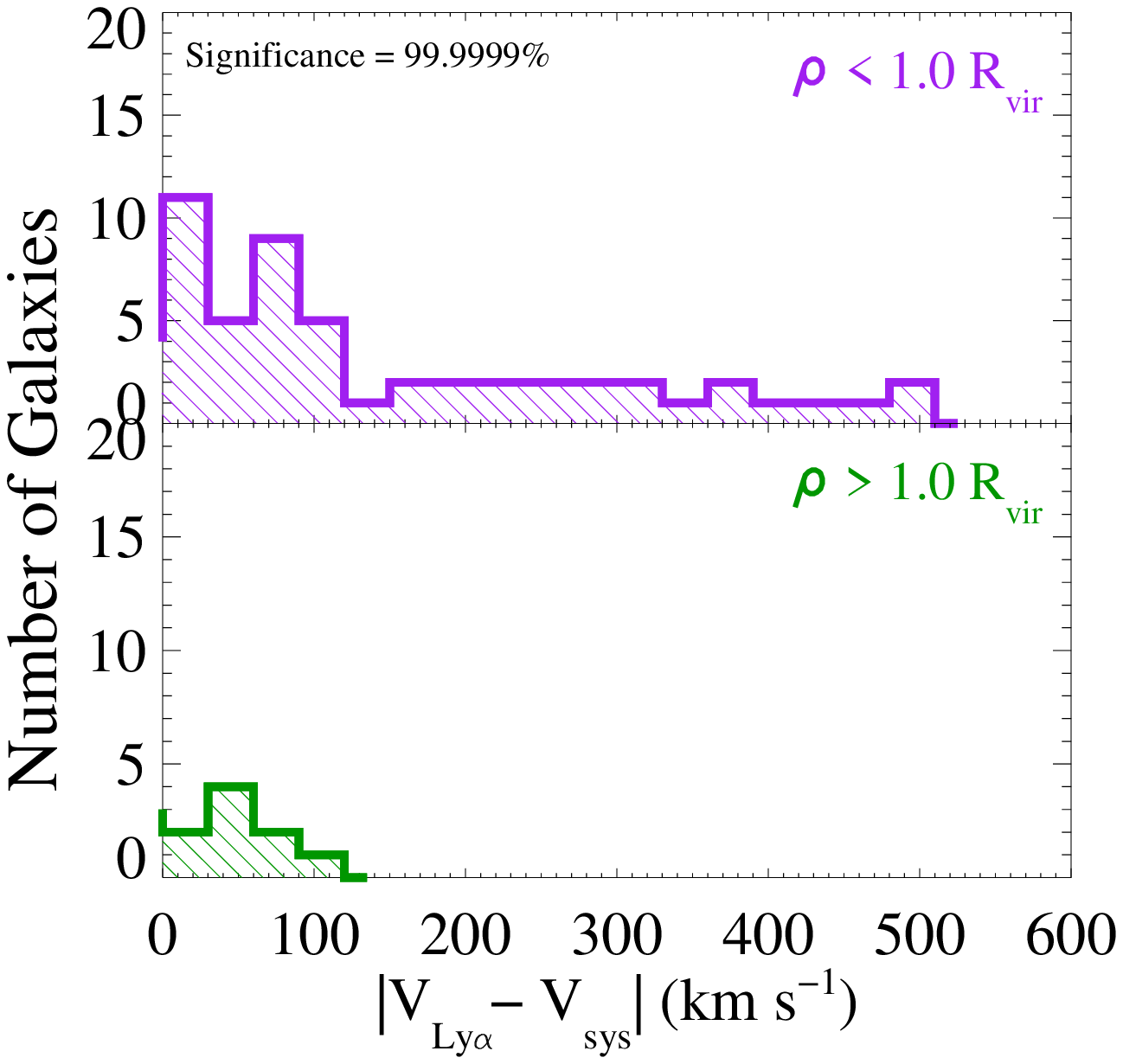}    
\includegraphics[trim = 0mm 0mm 35mm 0mm, clip,scale=0.65,angle=-0]{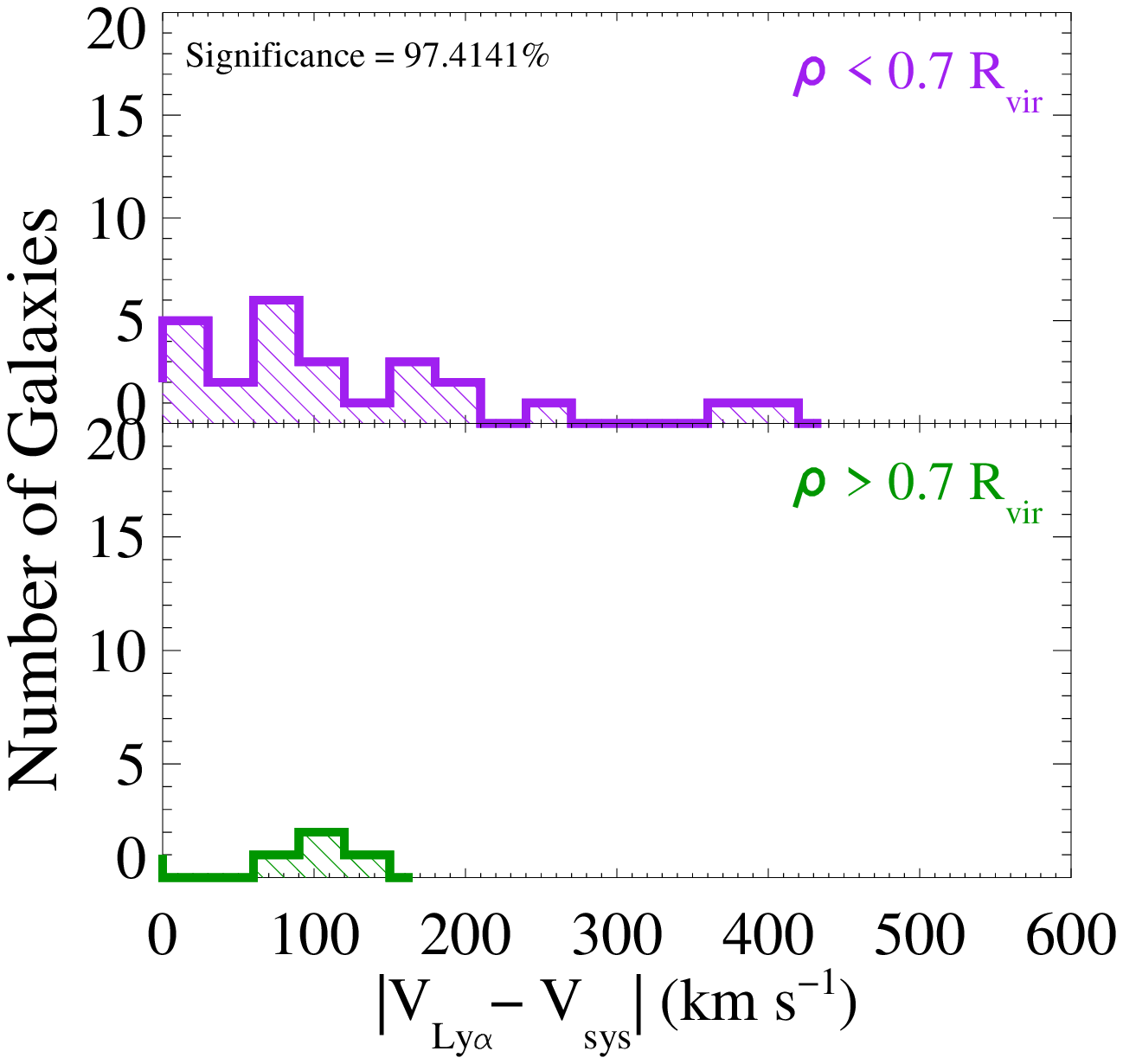}    
\caption{ Left: Histogram showing the distribution of  $\rm |V_{Ly\alpha}-V_{sys}|$ for the sightlines probing the CGM of blue galaxies within R$\rm _{vir}$ and outside of R$\rm _{vir}$. The ANOVA F-statistic test finds the two samples, the inner vs. the outer CGM, to have different \Lya centroid velocities at a 99.99\% confidence Right: Histogram showing the distribution of  $\rm |V_{Ly\alpha}-V_{sys}|$ for the sightlines probing the CGM of red galaxies within 0.7~R$\rm _{vir}$ and outside of 0.7~R$\rm _{vir}$. The ANOVA F-statistic test finds the two samples, the inner vs. the outer CGM, to have different \Lya centroid velocities at a 97.4\% confidence.}
 \label{fig-hist_del_rhoR} 
\end{figure*}

\begin{figure}

\hspace{-1cm}
\includegraphics[trim = 10mm 0mm 5mm 0mm, clip,scale=0.65,angle=-0]{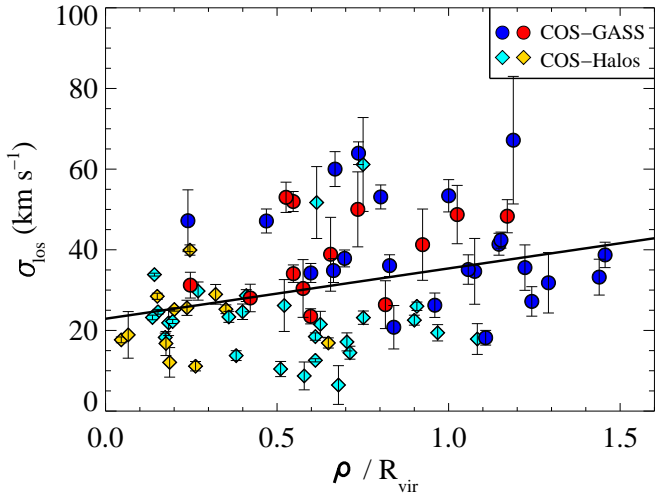}    
\caption{ The line-of-sight velocity dispersion of the strongest component of the \Lya absorption line plotted as a function of normalized impact parameter. The Kendall''s $\tau$ test indicates a correlation that is significant at the 99.6\% confidence level.}  
 \label{fig-sigma_normrho} 
\end{figure}

\begin{figure*}
\includegraphics[trim = 15mm 0mm 0mm 0mm, clip,scale=0.6, angle=-0]{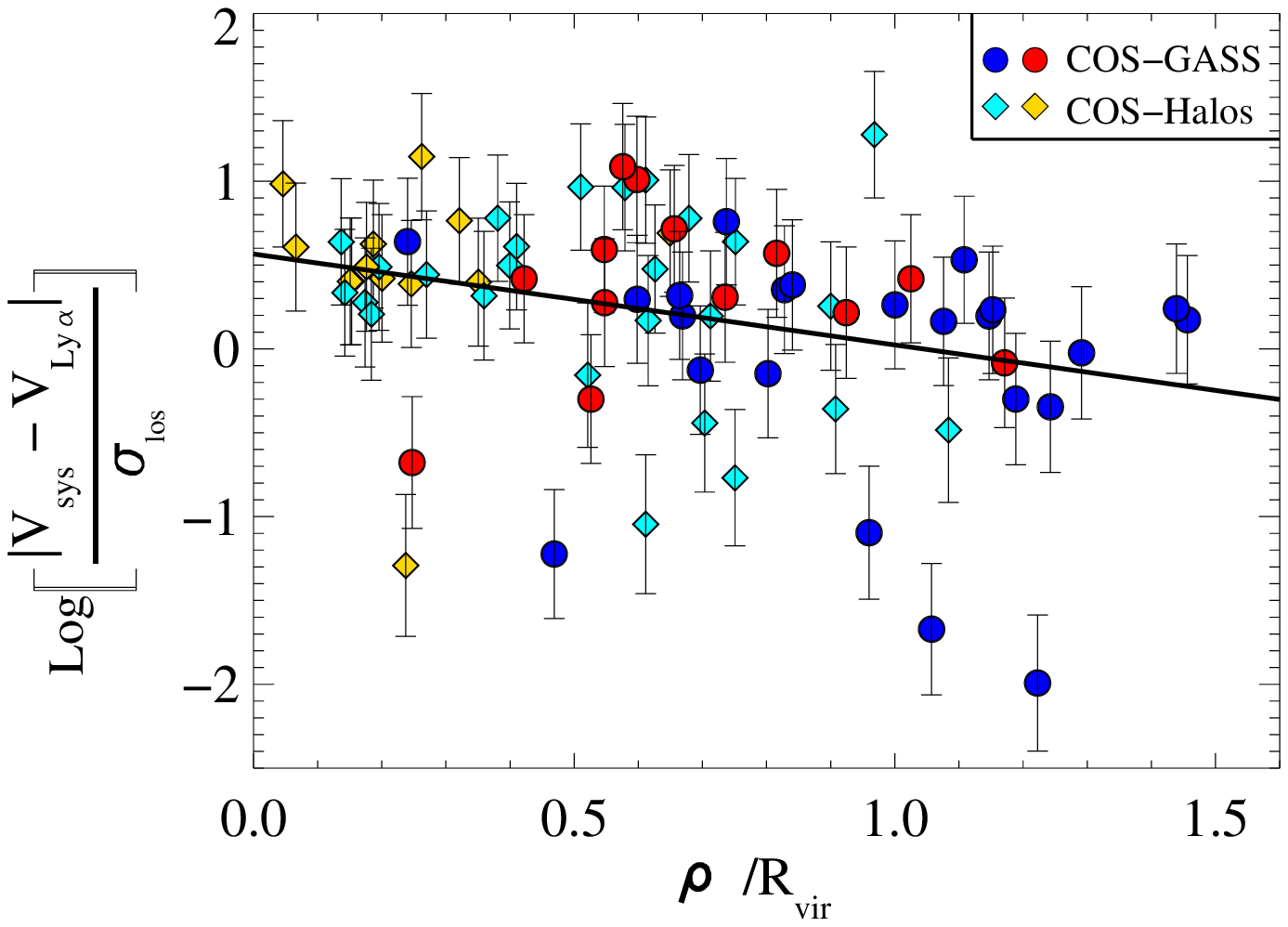}   
\includegraphics[trim = 15mm 0mm 0mm 0mm, clip,scale=0.6, angle=-0]{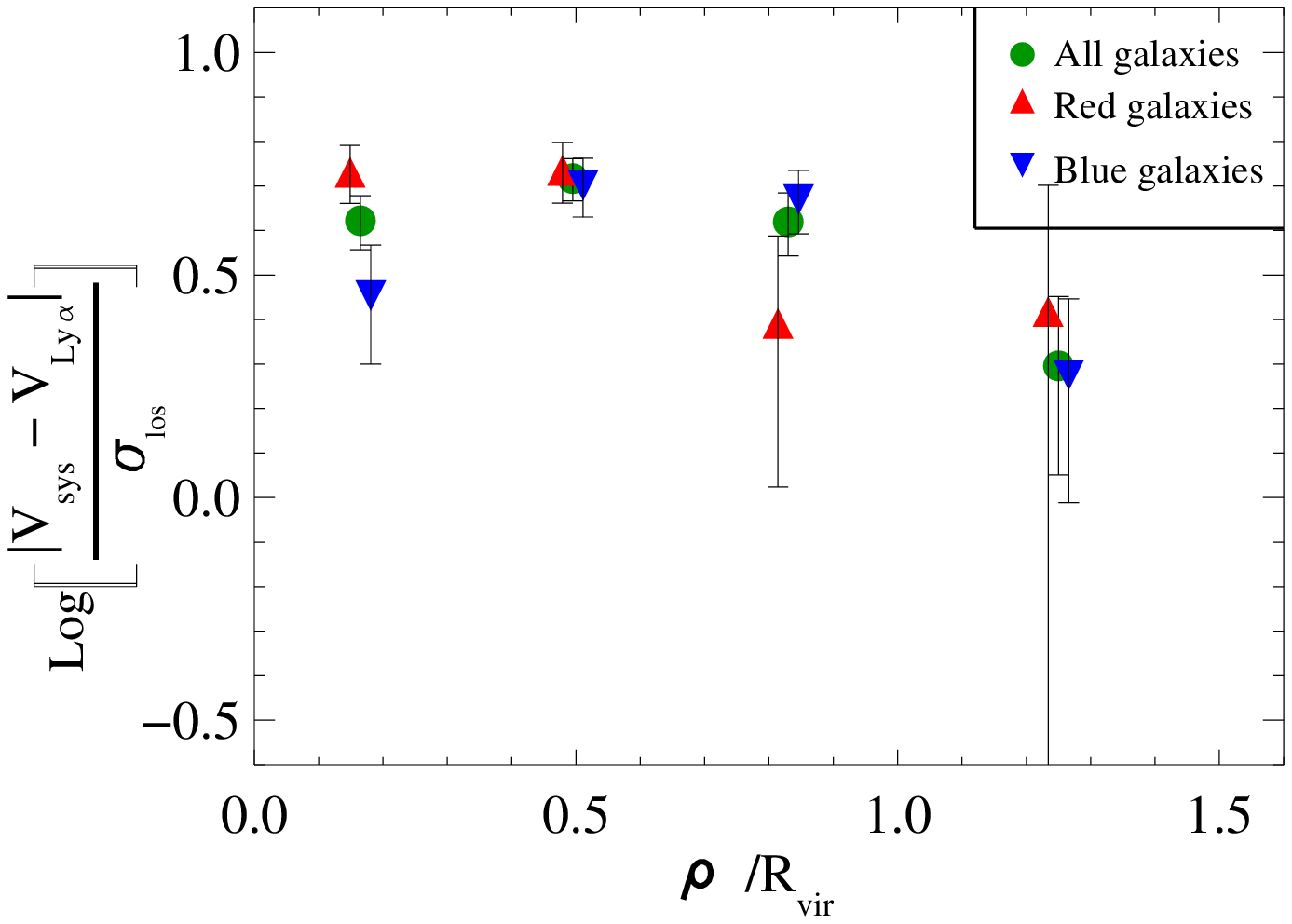}   
\caption{The log of the ratio of $\Delta v$ (the velocity difference between the \Lya absorption line and galaxy systemic velocity) and $\sigma_{los}$ (the line-of-sight velocity dispersion) as a function of normalized impact parameter. The Kendall's $\tau$ test indicates that the correlation is significant at the 99.9\% confidence level. The solid line is the best fit. The bottom panel shows the same relation but using the mean values in bins of normalized impact parameter. The drop in the ratio in the outer CGM is evident.}
\label{fig-vel2_rho_Rvir} 
\end{figure*}

 \begin{figure*}
\includegraphics[trim = 5mm 0mm 0mm 0mm, clip,scale=0.535,angle=-0]{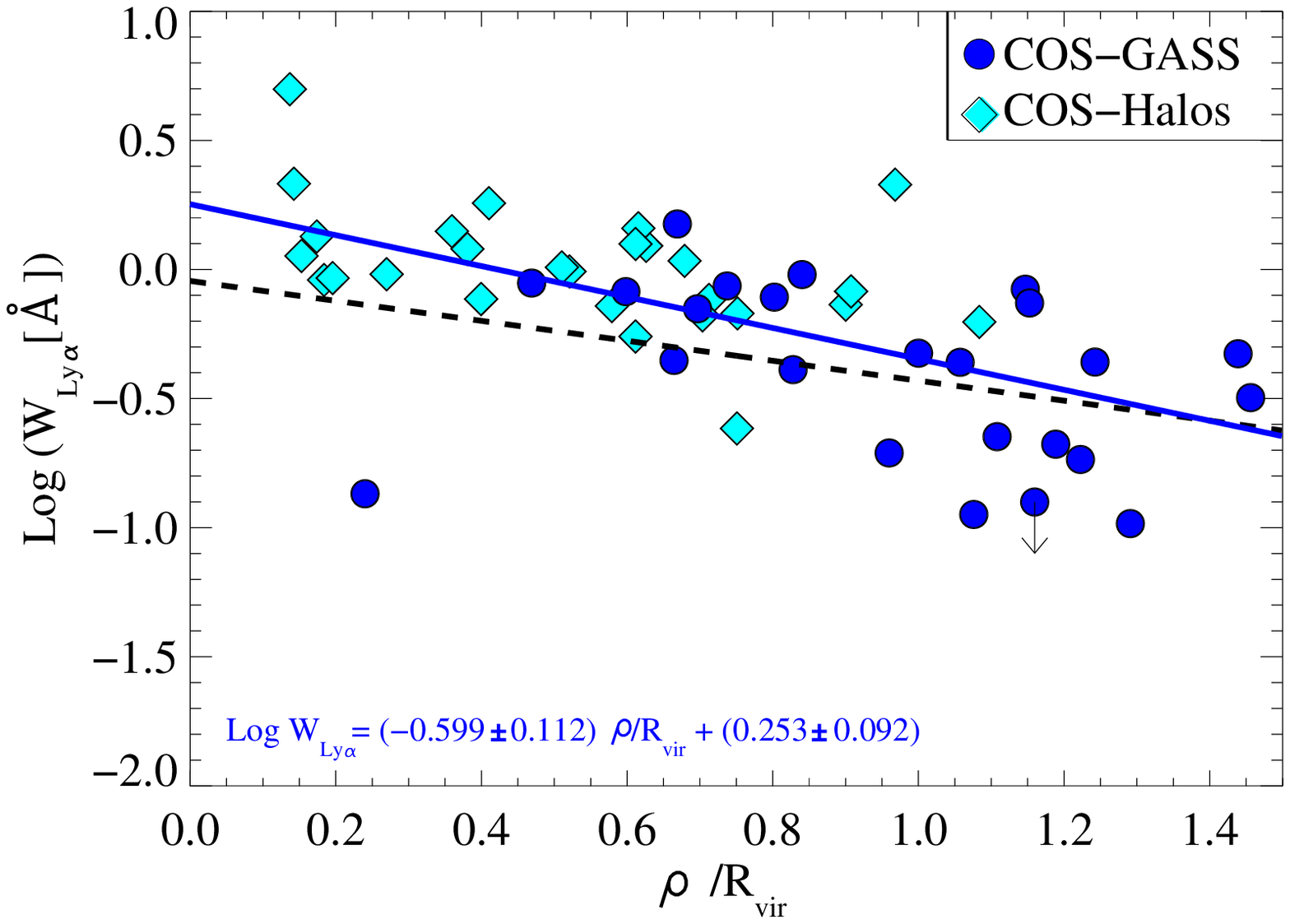} 
\includegraphics[trim = 5mm 0mm 0mm 0mm, clip,scale=0.535,angle=-0]{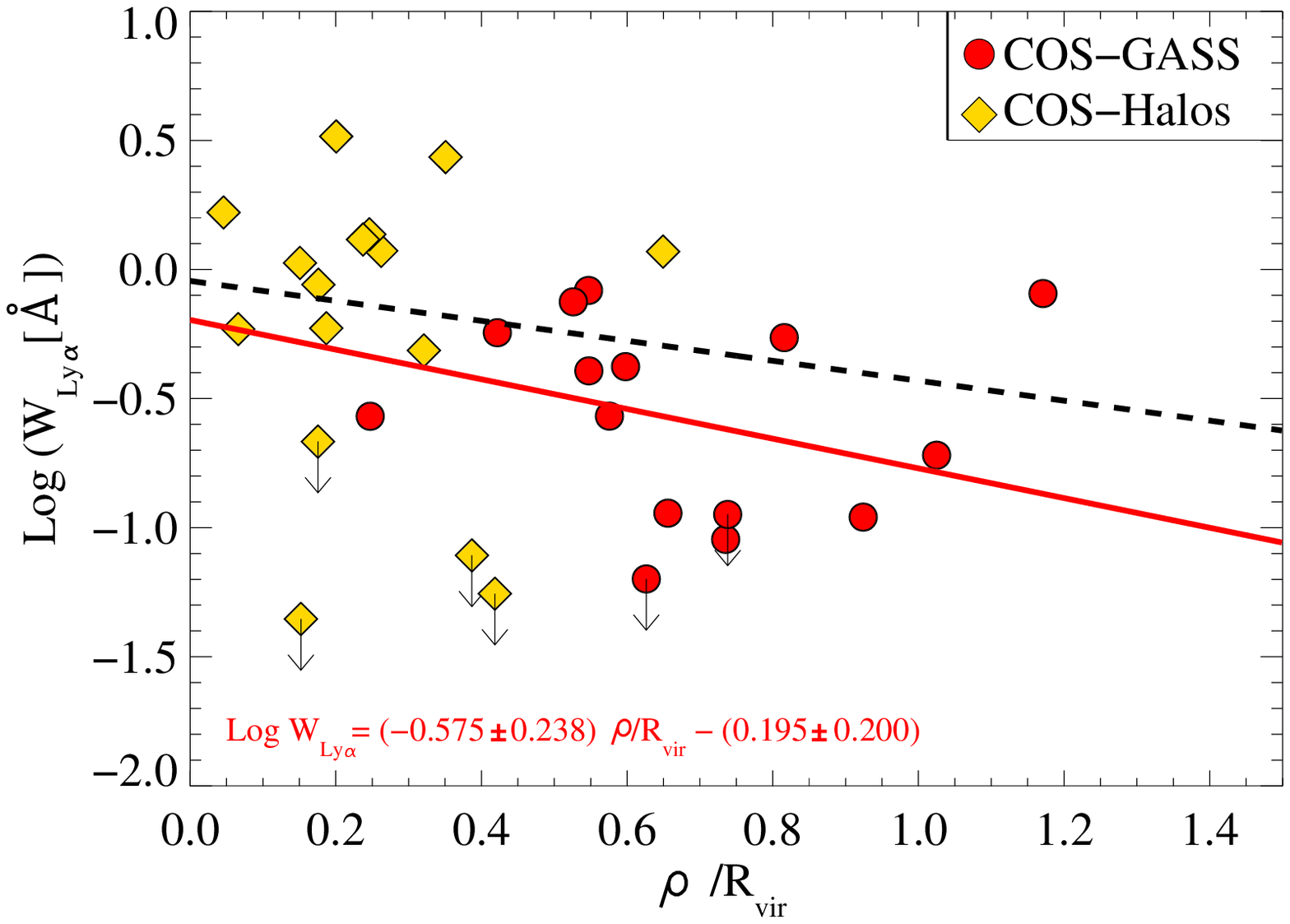}     
\caption{Variation of \Lya equivalent width as a function of normalized impact parameter i.e. $\rm \rho/R_{vir}$.  The galaxies are divided into two classes, blue galaxies (left panels) and red galaxies (right panels), based on their sSFR being above or below $\rm 10^{-11}~M_{\odot}~yr^{-1}$. The solid blue and red lines show the best-fit to the plotted data using the Buckley-James method. The fits were performed using the survival analysis software ASURV that takes into account the censored data. The dashed black line denotes the fit to the entire data set as shown in left panel of fig-\ref{fig-W_rho_Rvir}.   The parameters describing the best-fit lines are printed at the bottom left corner.}
 \label{fig-lya_rho_red_blue} 
\end{figure*}

 \begin{figure}
\vspace{0.5cm}
\includegraphics[trim = 20mm 0mm 0mm 0mm, clip,scale=0.6,angle=-0]{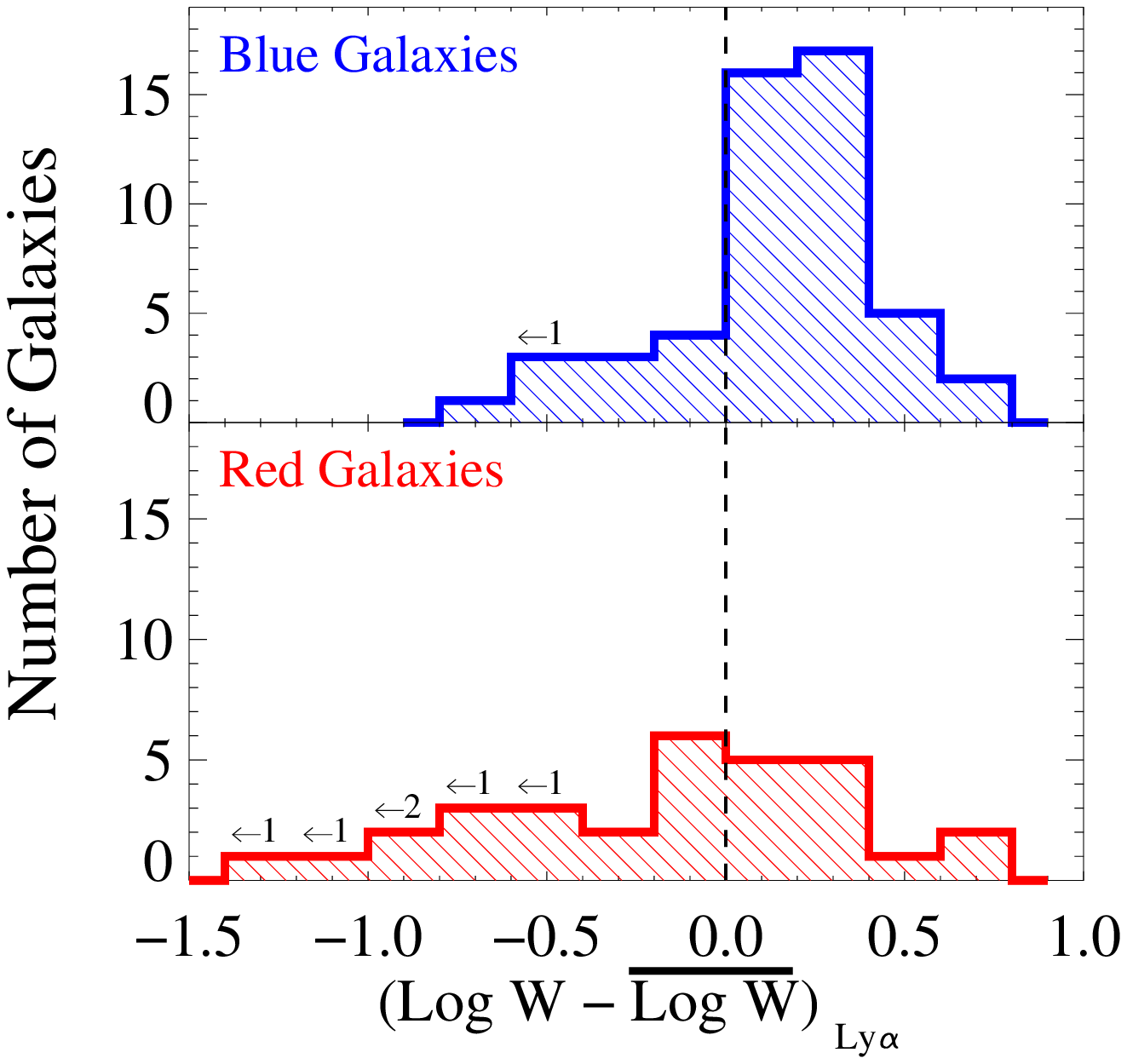} 
\caption{Distribution of [Log W - $\rm \overline{Log~W}]_{Ly\alpha}$ i.e. the offset in the \Lya equivalent width from the best-fit line described in Figure~3 for the combined samples of blue and red galaxies. The black arrows refer to the limiting cases included in each bin.  The Logrank test taking into account the censored data finds the two samples, blue galaxies and red galaxies, to differ in their \Lya equivalent width at a confidence level of 99.9\%.}
 \label{fig-his_lya} 
\end{figure}

 \begin{figure*}
\includegraphics[trim = 5mm 0mm 0mm 0mm, clip,scale=0.535,angle=-0]{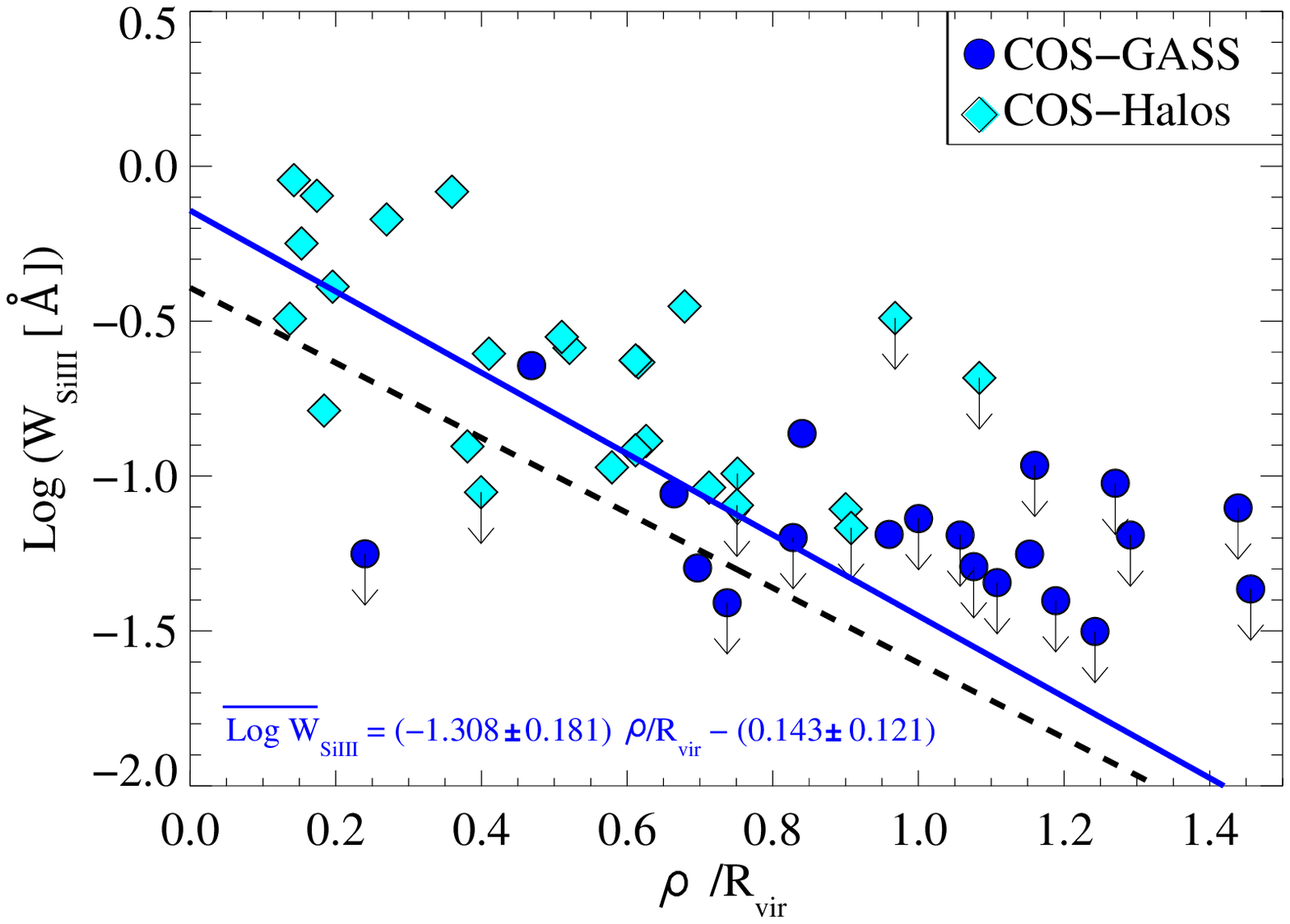} 
\includegraphics[trim = 5mm 0mm 0mm 0mm, clip,scale=0.535,angle=-0]{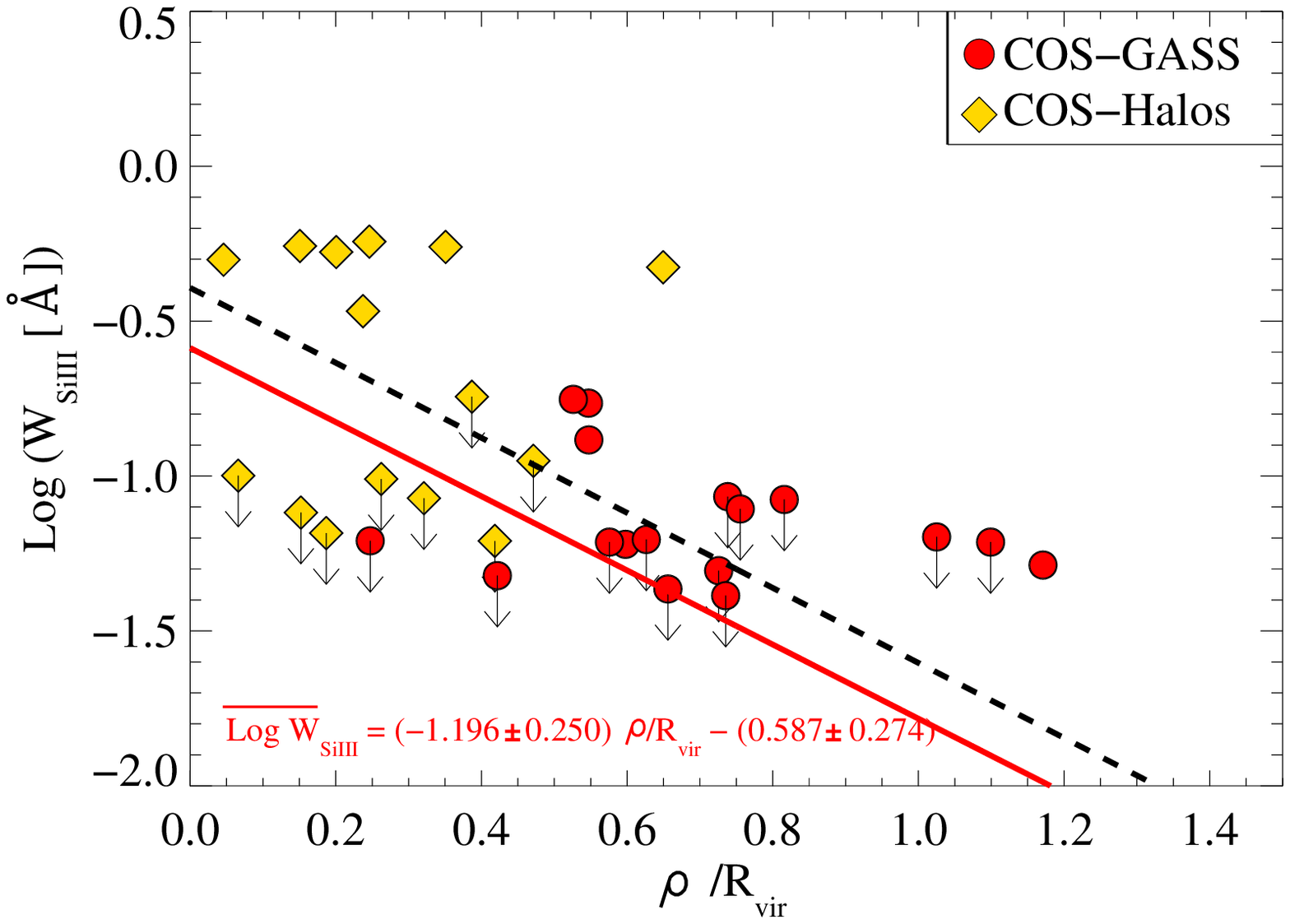}     
\caption{Variation of Si~III equivalent width with  normalized impact parameter i.e. $\rm \rho/R_{vir}$.  The galaxies are divided into two classes, red and blue, depending on whether the sSFR less or greater than $\rm10^{-11}~M_{\odot}~yr^{-1}$. The solid  blue and red lines show the best-fit to the plotted data using the Buckley-James method. The fits were performed using the survival analysis software ASURV that takes into account the censored data. The dashed black line denotes the fit to the entire data set as shown in the right panel of fig-\ref{fig-W_rho_Rvir}.   The parameters describing the best-fit lines are printed at the bottom left corner. }
 \label{fig-Si3_rho_red_blue} 
\end{figure*}

 \begin{figure}
\vspace{0.5cm}
\includegraphics[trim = 20mm 0mm 0mm 0mm, clip,scale=0.6,angle=-0]{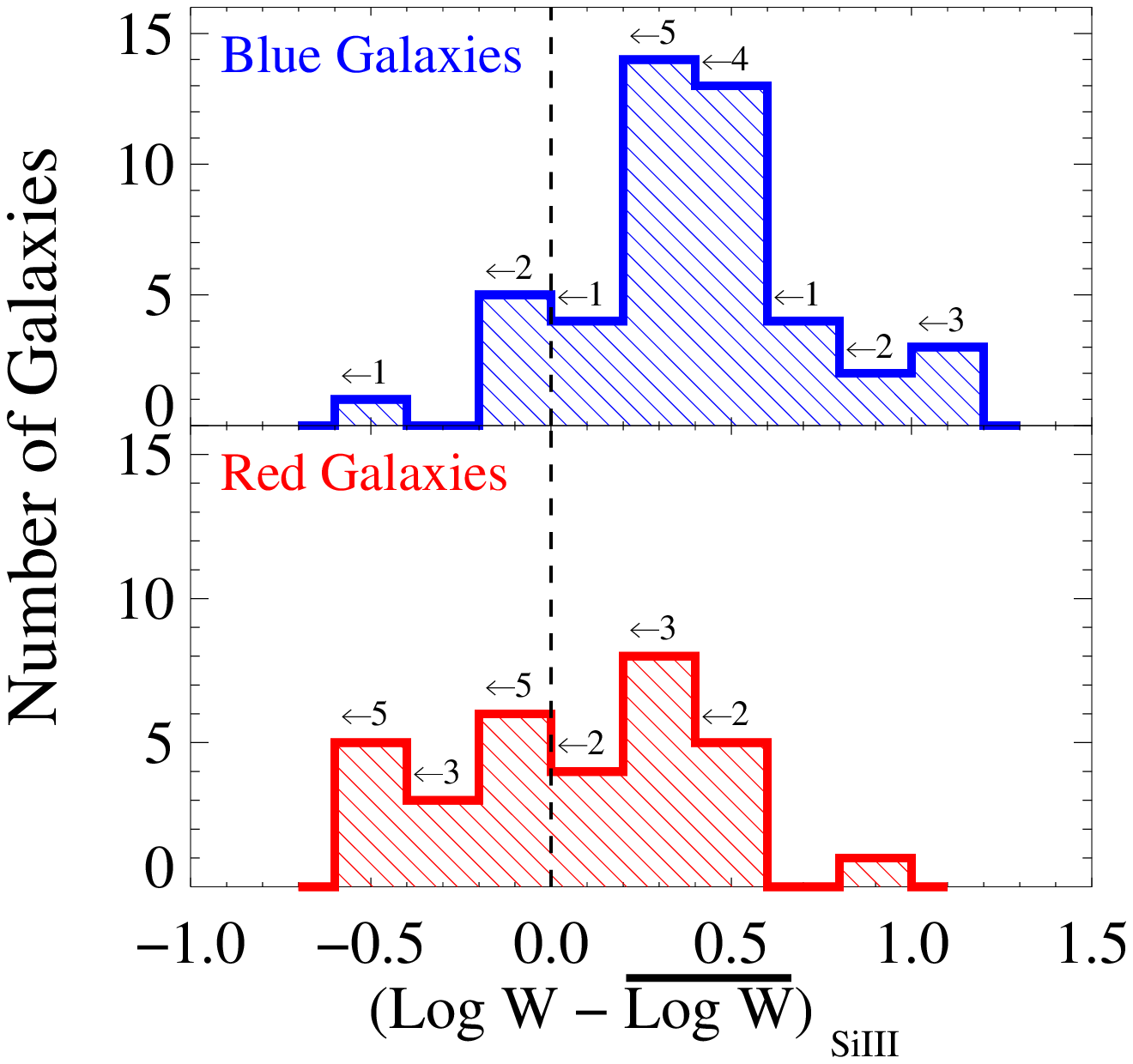} 
\caption{Distribution of [Log W - $\rm \overline{Log~W}]_{SiIII1206}$ i.e. the offset in the Si~III equivalent width from the best-fit line described in Figure~3 for the blue and the red galaxies. The black arrows refers to the limiting cases included in each bin. The Logrank test taking into account the censored data  finds the two samples, blue galaxies and red galaxies, to differ in their Si~III equivalent width at sightly more than 99.8\% confidence. }
 \label{fig-his_Si3} 
\end{figure}

\begin{figure}
\vspace{0.5cm}
\includegraphics[trim = 0mm 0mm 0mm 0mm, clip,scale=0.61,angle=-0]{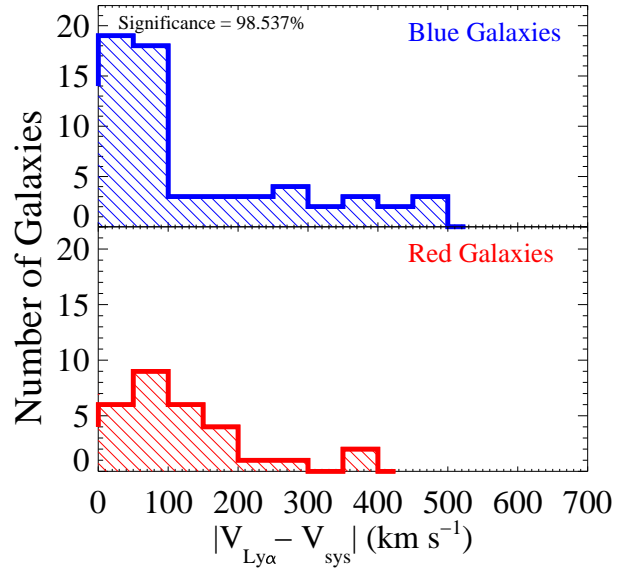}  
\caption{Histogram showing the distribution of  $\rm |V_{Ly\alpha}-V_{sys}|$ for the blue and the red galaxies. The ANOVA F-statistic test finds the two samples, blue galaxies and red galaxies, to have different \Lya centroid velocity distributions at a 98.5\% confidence level. }
 \label{fig-hist_del} 
\end{figure}

\begin{figure*}
\includegraphics[trim = 5mm 0mm 0mm 0mm, clip,scale=0.535,angle=-0]{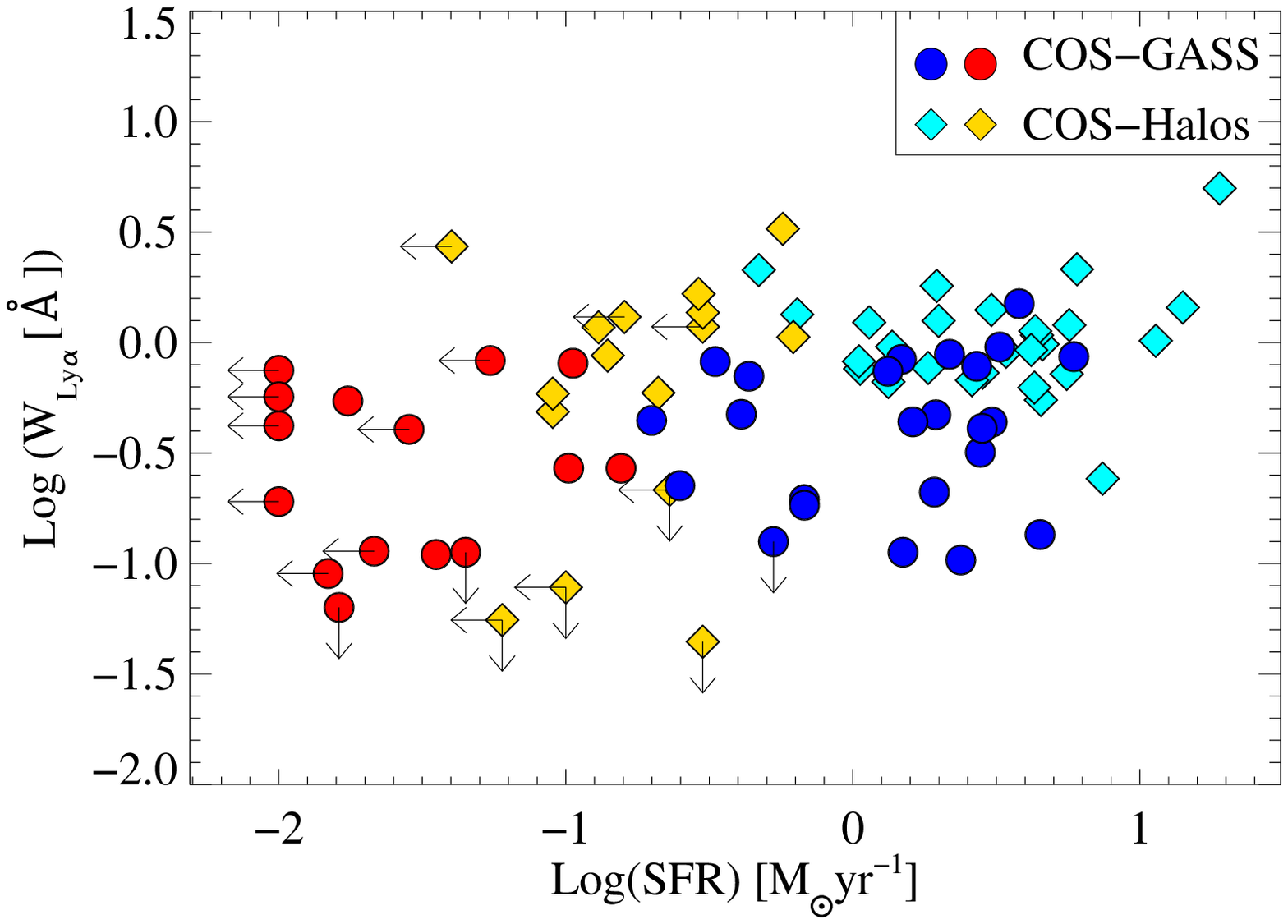} 
\includegraphics[trim = 5mm 0mm 0mm 0mm, clip,scale=0.535,angle=-0]{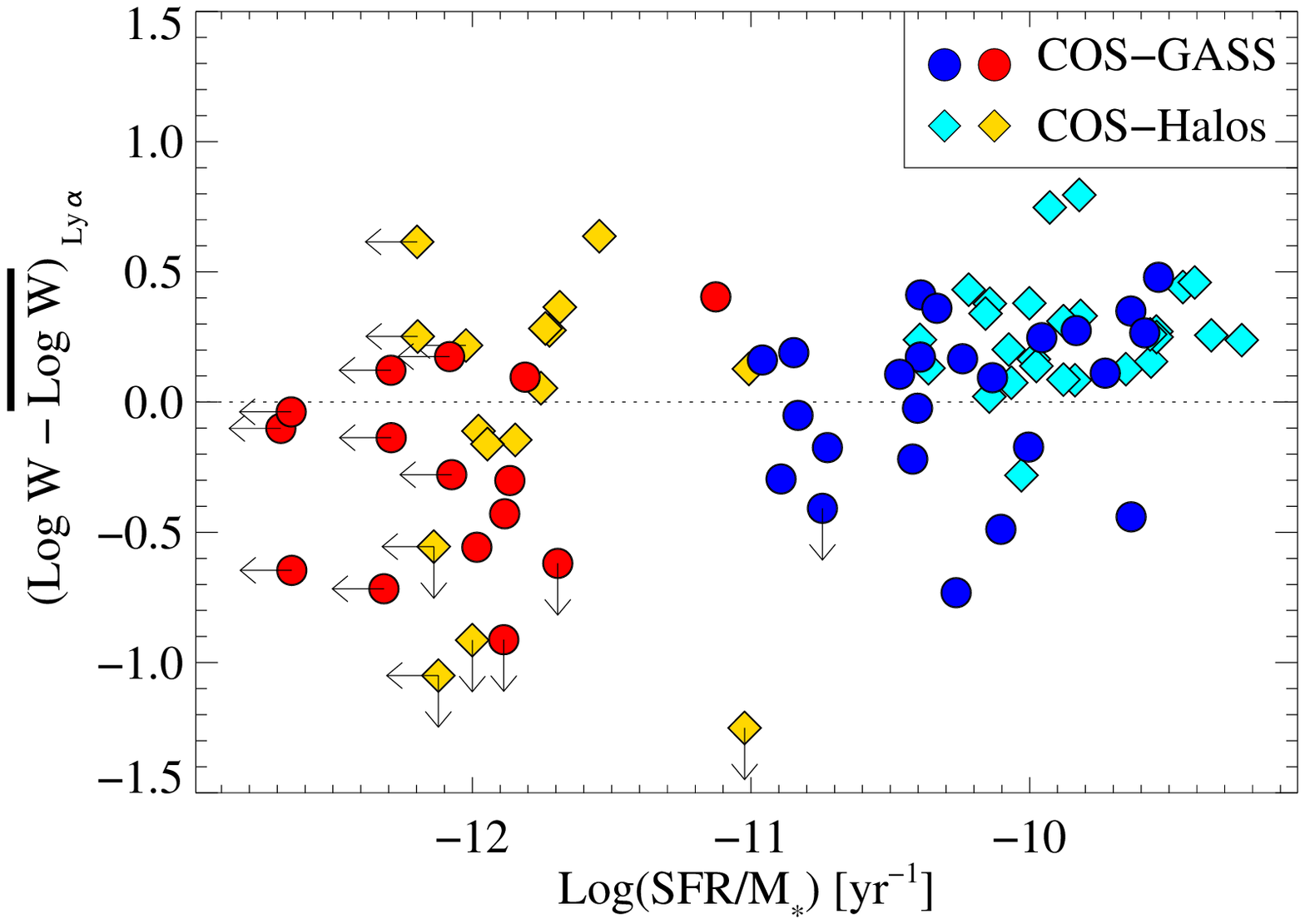}     
\caption{LEFT: \Lya equivalent width as a function of star formation rate (SFR) of the host galaxy. RIGHT: Impact parameter corrected equivalent width ([Log W - $\rm \overline{Log~W}]_{Ly\alpha}$) as a function of specific SFR of the galaxies. Excess equivalent width is defined as the ratio of the observed \Lya equivalent width and that predicted by the best fit line for the entire sample as shown in Figure~\ref{fig-W_rho_Rvir}. The correlation between the excess in equivalent width of \Lya and sSFR is measured at the 99.99\% confidence level using survival analysis code ASURV. }
 \label{fig-lya_sfr} 
\end{figure*}

\begin{figure*}
\includegraphics[trim = 5mm 0mm 0mm 0mm, clip,scale=0.535,angle=-0]{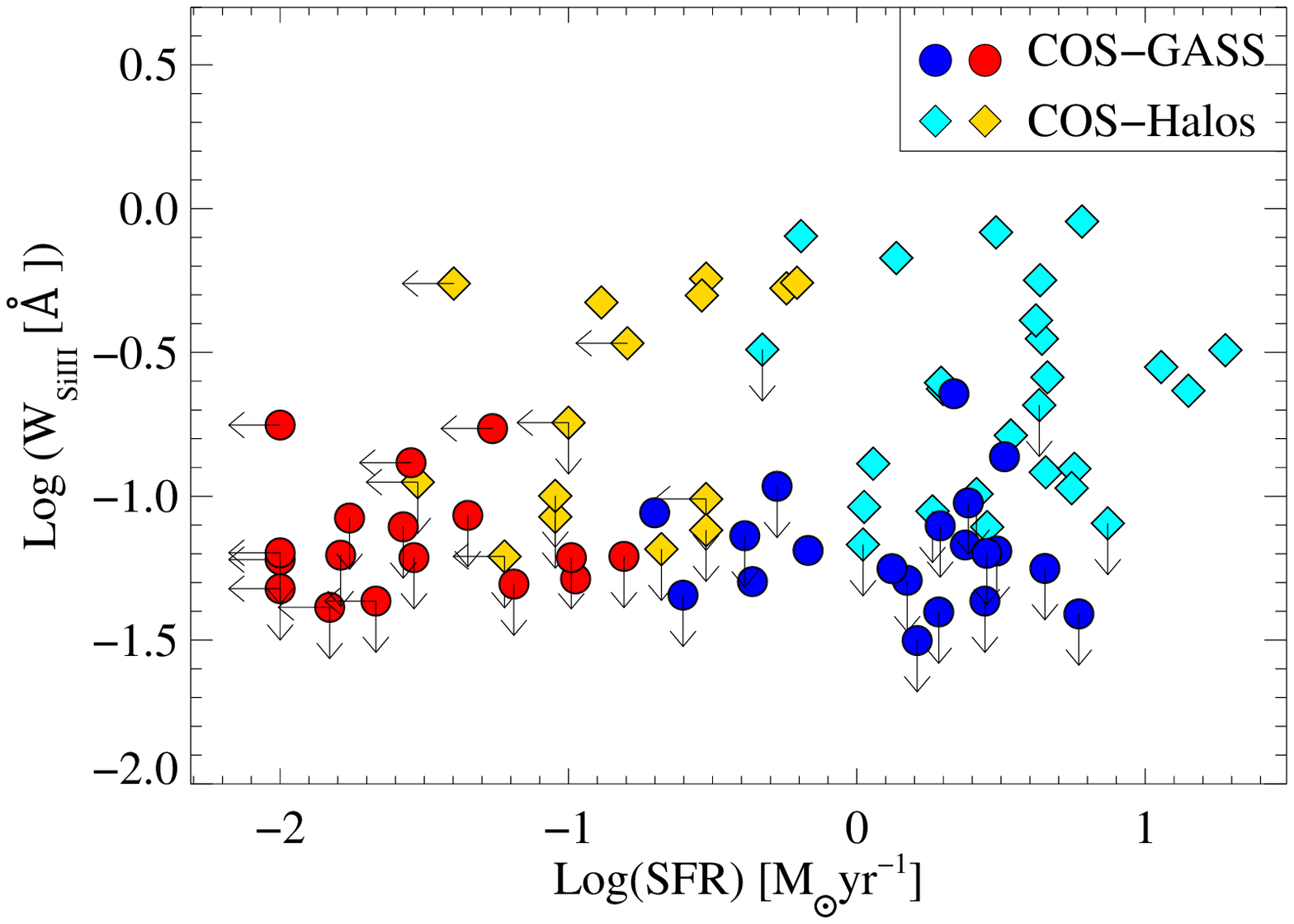} 
\includegraphics[trim = 5mm 0mm 0mm 0mm, clip,scale=0.535,angle=-0]{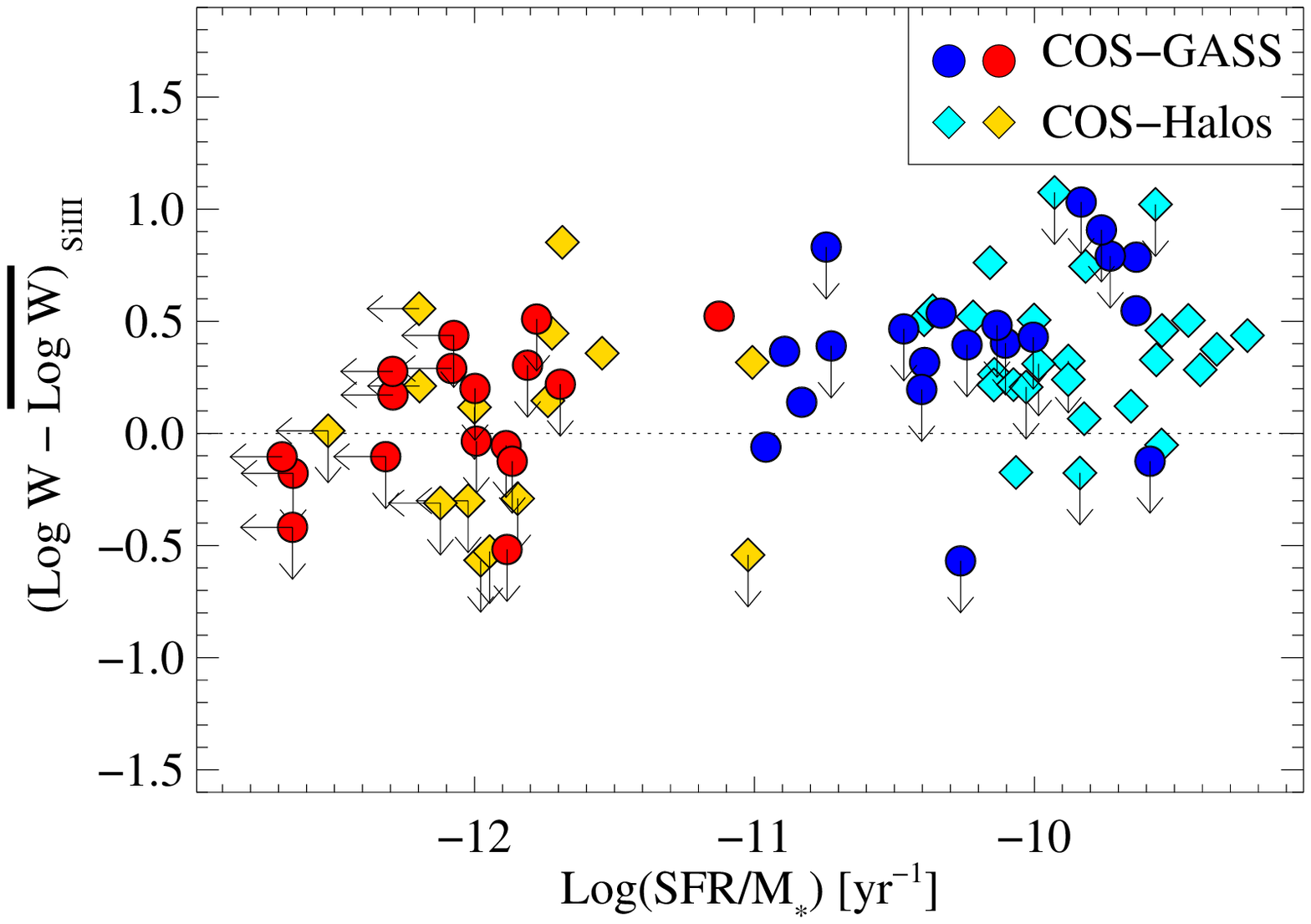}    
\caption{LEFT: Si~III equivalent width as a function of star formation rate (SFR) of the host galaxy. RIGHT: The excess in  Si~III equivalent width ([Log W - $\rm \overline{Log~W}]_{SiIII1206}$)  as a function of specific SFR of the galaxies. The excess in Si~III equivalent width represents the difference between the observed  Si~III equivalent width and that predicted by the best fit line as shown in Figure~\ref{fig-W_rho_Rvir}. The correlation between the excess in equivalent width of Si~III and sSFR is measured at the 99.7\% confidence level using survival analysis code ASURV.}
 \label{fig-Si3_sfr} 
\end{figure*}

\subsection{An Overview of the CGM Properties}
 
We begin by summarizing the basic structural and kinematic properties of the CGM. Later, we will consider the dependence of these properties on the star forming characteristics of the central galaxy.

\subsubsection{Structure}

The dark matter halo mass of the galaxy should influence the size and kinematic properties of the CGM \citep[][and references therein]{ford16,hummels12}. For example, it is expected that a galaxy with a larger halo mass could contain a more massive CGM and gravitationally bind it to larger radii \citep{chen01b}. In order to explore the radial profile of the CGM while accounting for the large range in halo mass, we use the variable $\rho/\rm R_{vir}$, which we refer to as the normalized impact parameter \citep[e.g.][]{stocke13}. This parameter scales the impact parameter ($\rho$) in terms of the size of the dark matter halo ($\rm R_{vir}$). By doing so, we standardize the position of the sightlines for galaxies of different halo masses and consequently CGM sizes. Similar analyses have been performed on different data sets by \citet{stocke13} and \citet{liang14}, and on COS-Halos and COS-GASS by \citet{tumlinson13} and \citet{borthakur15} respectively. 

We show the radial distribution of the equivalent width of \Lya normalized with respect to the virial radius of the galaxies in Figure~\ref{fig-W_rho_Rvir}. The distribution can be fit as a exponential with a scale-length of $\rm 1.1~R_{vir}$, i.e. $\rm W_{Ly\alpha} = A~e^{-\rho/1.1R_{vir}}~\AA$, where the normalization factor, A, is equal to 0.9 $\rm \AA$. The fit was derived using the Buckley-James \footnote{The Buckley-James method is a semi-parametric regression method} method \citep{buckley_james_79} and Expectation-maximization algorithm as implemented in the survival analysis software package, ASURV \citep{ASURV}. The equivalent width data presented here are the same as that of Figure~2 of \citet{borthakur15}, however, the abscissa is different as we have adopted the \citet{kravtsov14,liang14} formalism with modifications based on the findings of \citet{mandelbaum16} for halo masses and virial radii. In addition, the fit presented here takes into account the censored data and hence has slightly different parameters.


Similarly, the radial distribution of the equivalent width of \ion{Si}{3} (see right panel of   Figure~\ref{fig-W_rho_Rvir}) can be fit as a exponential with a scale-length of 0.4 $\rm R_{vir}$ i.e. $\rm W_{SiIII} = 0.4~ e^{-\rho/0.4R_{vir}} ~ \AA$. Almost all of our \ion{Si}{3} absorbers were detected inwards of 0.8$\rm R_{vir}$. The smaller characteristic size scale for the \ion{Si}{3} absorbers compared to \Lya and the lack of \ion{Si}{3} detections beyond about 0.8 $\rm R_{vir}$ are consistent with the conclusions drawn by \citet{liang14}.

Given the systematic radial decline in the strengths of both the \Lya and \ion{Si}{3} absorbers, we define a new parameter: the impact parameter corrected equivalent width (hereafter, the excess equivalent width, [Log W - $\rm \overline{Log~W}]_{ion}$). This refers to the offset in log $\rm W_{\rm ion}$ in any individual sightline with respect to the best fit exponential for the entire sample. This parameter is then independent of impact parameter biases and allows us to compare all the absorbers in a uniform way.

\subsubsection{Kinematics}

We turn now to the kinematic properties of the CGM. To begin, a useful way of visualizing these properties is via the one-dimensional cross-correlation function. The cross-correlation function between galaxies and velocity centroids of \Lya absorbers is presented in the Figure~\ref{fig-cc_lya}. The cross-correlation function shows a strong signal for the presence of \Lya absorbers within 120~\kms of the galaxy systemic velocity. The systemic velocity is defined as the velocity corresponding to the optical spectroscopic redshift from SDSS that traces the stars that form the bulk of the baryonic material in the central region of the galaxies. 

The uncertainties were derived using a Jackknife error estimator and are indicated as the brown line. They are dominated by small number statistics although the random pairs were generated by using 10,000 Monte Carlo samples. The cross-correlation function also takes into account the same data analysis criteria as the observations. For example, we have taken into account the observational aspect of identifying features by designating absorbers within $\pm$600~\kms of the galaxy systemic as associated absorbers. For estimating the cross-correlation function, we randomly distributed the \Lya absorbers within $\pm$600~\kms of the galaxy systemic velocity. Therefore, it is not advisable to compare this analysis with blind galaxy-absorber cross-correlation functions such as those published by \citet{lanzetta98,chen05, ryan-weber06, wilman07, chen09, tejos12} and others.  We also considered each component of the \Lya absorption features as an individual absorber. This should be given due consideration when comparing our results to those from other studies that use spectrographs with significantly different velocity resolution from COS. 

Similar conclusions can be drawn for the \ion{Si}{3} transition. The distribution of \ion{Si}{3} is also closely related to that of \Lya absorption, although not all \Lya absorbers have associated \ion{Si}{3} absorption. To illustrate the velocity distribution of \ion{Si}{3} with respect to \Lya, we calculated the cross-correlation function between the velocity offset between \Lya and \ion{Si}{3}. In doing so, we preserved the relationship between distribution of \Lya centroids w.r.t the galaxy systemic - which implies that \Lya is not randomly distributed w.r.t $v=0$. The cross-correlation function is presented in Figure~\ref{fig-cc_Si3}. The red line indicates the cross-correlation function with the uncertainties shown in brown. The uncertainties were calculated using the same procedure as described previously. The correlation is strongest within $\pm$40\kms of the \Lya absorbers and drops gradually. The signal-to-noise is greater than $\rm3.5\sigma$ up to 160 \kms.

We now consider how the CGM kinematics depend upon the halo mass. Figure~\ref{fig-vel_mass_esc} shows the velocity offset of the absorbers relative to the systemic velocity of the galaxy ($|v_{Lya} –-v_{sys}|$, hereafter $\Delta v$) plotted as a function of dark matter halo mass. We find that the velocity distribution of the centroids of majority of the \Lya absorbers (as well as other detected metal absorbers) to typically lie within 200~\kms of the systemic velocity of the galaxy. The escape velocities at impact parameters of 100~kpc, 200~kpc, 300~kpc, and R$\rm _{vir}$ are plotted as dashed and solid curves of different colors. The centroids of the strongest component are shown as the filled symbols and the other components are shown as open symbols. The colored vertical lines connecting the strongest component to the other components mark the full-width of the \Lya profiles. Nearly all the strongest \Lya components have velocity centroids within the escape velocity of their host galaxies. Therefore, we expect this material to be bound to the galaxies.

Another way to characterize the kinematics of the CGM is to use the widths of the absorption features rather than their velocity displacement. In Figure~\ref{fig-sigma_mhalo} we plot the distribution of line-of-sight velocity dispersion of the strongest component of the \Lya feature ($\sigma_{los} = b/ \sqrt{2}$, where $b$ is the Doppler parameter in our fits). Several features are noteworthy in this pair of figures. First, the lines are generally very narrow (mean $\sigma_{los} \sim$ 30 \kms). Second, there is no trend for these widths to increase as the halo mass (virial velocity) increases (as was also the case for $\Delta v$). 

A compact representation of the information in the figures above is given in Figure~\ref{fig-vel_mass}. We define the kinematic parameter $W = (\sigma_{bins}^2 + \sigma_{avg}^2)^{1/2}$. Here, $\sigma_{bins}$ is the velocity dispersion of the distribution of the velocity differences between the \Lya absorber and the galaxy systemic velocities within a given bin in $\rm log~M_{halo}$. The term $\sigma_{avg}$ is the average value of $\sigma_{los}$ of the individual \Lya absorbers in this same bin in halo mass. Again, we see no dependence of CGM kinematics on halo mass. In particular, the low (sub-virial) velocity spread of CGM absorbers in the halos of massive red galaxies has been noted before \citep{zhu14,huang16}. We will discuss the possible implications of these results in section~4.

We can also examine the radial dependences of the CGM kinematics. In Figure~\ref{fig-vel_rho_Rvir} we plot the \Lya velocity distribution as a function of normalized impact parameter. This figure suggests that the velocity offset of the absorbers from systemic velocity ($\Delta v$) drops in the outer CGM. The differing kinematic properties of the inner {\it vs.} the outer CGM are shown in histogram form in Figure~\ref{fig-hist_del_rhoR}. An F-test shows that the distributions differ at $>$99.99\% (97.4\%) confidence level for the blue (red) galaxies. In Figure~\ref{fig-sigma_normrho} we show a similar plot of the radial dependence of the line-of-sight velocity dispersions ($\sigma_{los}$) of the \Lya absorbers. This shows that there is a statistically significant (99.6\% confidence level) trend for $\sigma_{los}$ to {\it increase} with increasing impact parameter.

We next define the dimensionless quantity $\Delta v/\sigma_{los}$, and plot it in Figure~\ref{fig-vel_rho_Rvir} as a function of normalized impact parameter. The most interesting result is that $\Delta v/\sigma_{los}$ has a mean value of about four interior to the virial radius. This figure also shows that the ratio declines in the outer CGM (a result significant at the 99.9\% confidence level). This is due to the combined effects of the drop in the $\Delta v$ in the outer CGM and the radial rise in $\sigma_{los}$ that were described above.

\subsection{The CGM in Blue {\it vs.} Red Galaxies}

We now compare the radial distributions of the absorbers in the CGM surrounding the blue {\it vs.} the red galaxies. The radial \Lya profiles as a function of normalized impact parameter ($\rho/\rm R_{vir}$) are shown in Figure~\ref{fig-lya_rho_red_blue}. One clear difference is the dispersion in the data between the two sub-samples. The blue galaxies show a fairly uniform radial distribution with a $>$95\% detection rate of \Lya absorbers out to $\sim \rm R_{vir}$. On the other hand, the red galaxies show a much larger dispersion in the radial distribution (as indicated by weak absorption features as well as non-detections with very good upper limits). It is worth noting that red galaxies do occasionally exhibit strong \Lya absorbers associated with their CGM, but their detection rate is not as large as in blue galaxies. This is particularly true for the inner CGM, and suggests that the warm CGM is more patchy in the red galaxies (has a smaller areal covering factor). 

The dashed black line is the fit to entire sample, whereas the blue and red solid lines are fits to blue and red galaxies respectively. The implied exponential scale-lengths for the blue and red galaxies are similar to one another (0.65 $\rm R_{vir}$ and 0.75 $\rm R_{vir}$ respectively). The difference in the normalization of the profiles of 0.45 dex (blue vs. red) reflects the patchy nature of the absorbers in the CGM of the red galaxies.

Histogram representations of the excess \Lya equivalent widths for the blue and red galaxies are presented in Figure~\ref{fig-his_lya}. The dispersion in the red galaxy sub-sample is much higher than in the blue galaxy sub-sample. Again, this signifies the difference in the covering fraction of neutral gas between the two populations. The distribution in the excess \Lya equivalent widths between the blue and red galaxies is significant such that these two sub-samples can be considered as different populations with 99.9\% confidence based on Logrank test statistics using the software package ASURV.

We find similar results for \ion{Si}{3} (Figure~\ref{fig-Si3_rho_red_blue}). The exponential scale lengths are similar for the blue and red galaxies (0.33 $\rm R_{vir}$ and 0.36 $\rm R_{vir}$ respectively). The normalization of the fit to the equivalent width radial distribution is 0.44 dex higher for the blue galaxies, which again reflects the patchy nature of the absorbers in the CGM of the red galaxies. The histogram showing the distributions of the excess \ion{Si}{3} equivalent widths is shown in Figure~\ref{fig-his_Si3}.  The difference in the distribution of the excess \ion{Si}{3} equivalent width between the red and the blue galaxies is significant at the 99.8\% confidence level. 

By and large, the kinematic properties of the CGM are quite similar between the blue and red galaxies. However, one notable difference is highlighted in Figure~\ref{fig-hist_del}, which shows histograms of $\Delta v$ for the individual \Lya absorbers. 
While in both samples, the majority of the values for $\Delta v$ are less than 100 \kms, the blue galaxy histogram has a pronounced tail extending out to $\Delta v \sim $500 \kms. An Analysis of Variance (ANOVA) F-test reveals that the blue and red samples differ at the 98.5\% confidence level.

\subsection{Correlation Between CGM Properties and SFR}

In the discussion above we have simply classified galaxies as star-forming (blue) or quiescent (red). In this section we focus on quantitative measures of the SFR and the sSFR. In our recent study, we found a strong correlation between the neutral hydrogen content in the interstellar medium of galaxies (traced by the 21~cm hyperfine transition) and the \Lya equivalent width in the outer CGM probed by the COS-GASS sample \citep{borthakur15}. We also found correlations between \Lya strengths and both SFR and sSFR, but these were significantly weaker than those with the \HI 21~cm mass or mass fraction. Here, we reinvestigate the correlations with SFR and sSFR using the much larger combined sample that better covers the full radial range of the CGM. 
 
As seen in Figure~\ref{fig-lya_sfr}, we find a positive correlation between the equivalent width of the \Lya absorbers and the SFR at the 99.8\% confidence level. We find an even stronger correlation with the excess equivalent width [Log W - $\rm \overline{Log~W}]_{Ly\alpha}$ and the galaxy sSFR (at the 99.99\% confidence level\footnote{The test was performed on our censored data using the astronomy survival analysis code ASURV \citep{ASURV}.  ASURV is capable of handling single and doubly censored data. The accuracy of these probabilities can be affected by larger numbers of censored values and other conditions. Since less than a quarter of our sample has censored values, we do not expect substantial inaccuracies. However, caution is appropriate as is for results from Kendall's test  on any sample \citep{wang00}.}. We also find similar correlations between SFR and the equivalent width of the \ion{Si}{3} and between the excess \ion{Si}{3} equivalent width [Log W - $\rm \overline{Log~W}]_{SiIII1206}$ and sSFR (see Figure~\ref{fig-Si3_sfr}). 

\section{Discussion \label{sec:discussion}} 

We want to highlight a number of the results above and try to connect them together into a simple picture of the warm ionized phase of the CGM in both the blue (star-forming) and red (quiescent) galaxies. 

We have found that the distribution of \ion{Si}{3} absorbers is more compact than that of the \Lya absorbers for both the red and blue galaxies (with exponential length scales of $\sim$ 0.35 {\it vs.} 0.7 $\rm R_{vir}$ for the respective ions in both the blue and red galaxies). As a consequence, the detection fraction of \ion{Si}{3} for the full sample drops from 67\% inside 0.7 $\rm R_{vir}$ to only 17\% outside. This was also seen by \citet{liang14}, who interpreted it in terms of a boundary that represents the region that has been significantly enriched by metals expelled from the central galaxy (e.g. affected by stellar feedback at some point in the evolution of the galaxy and its CGM). We also see a change in the distribution of the velocity offsets between the \Lya lines and galaxy systemic velocities in the outer CGM. While this could be related to feedback processes, it could also be due to line-of-sight projection effects, if the flow pattern in the CGM has a strong radial component (inward and/or outward).  

Perhaps the most surprising result is that in neither the red nor blue populations do we see any trend for either the velocity offset of the \Lya absorbers with respect to the galaxy systemic velocity ($\Delta v$) or the line-of-sight velocity dispersions ($\sigma_{los}$) of the absorption lines to increase with increasing halo mass across a range of about 5 in implied virial velocity. The implied "sub-virial" velocities in the CGM around massive red galaxies have been noted before \citep{zhu14, huang16}, and are particularly mysterious. One possibility is that we are seeing the condensation of thermal instabilities out of a hot volume-filling phase \citep{voit15}. The hot gas is supported hydrostatically, and the denser cooler clouds that condense out have not (yet) been accelerated by gravity to the virial velocity of the halo. This would suggest that these clouds in massive red galaxies are transient (lifetimes less than a halo crossing time). 

Another possibility is that the flow of the absorbing material in the warm CGM of these galaxies is significantly affected by drag forces associated with a more massive hot volume-filling phase. In any case, it appears that the CGM dynamics are not purely determined by gravitational forces alone. 
Since we expect that the absorption-line systems we see are imbedded in a multi-phase halo, their dynamics is likely to be influenced by processes such as drag forces, thermal instabilities, turbulent mixing, merger dynamics, and feedback-driven outflows \citep{maller04, santillan07, kwak10, kwak11, joung12, joung12b, stinson12, ford14,  fielding16, suresh15}.

The role of drag forces in reducing the velocities of the CGM clouds in massive halos is particularly interesting to consider. Following \citet{bordoloi16}, it is straightforward to show that the terminal velocity for a CGM cloud that is significantly affected by drag is given by: 
\begin{equation}
v_{term} \sim v_{vir} (M_{cs}/M_{vf})^{1/2}
\end{equation}
Here $v_{vir}$ is the virial velocity of the halo, and $M_{vf}$ and $M_{cs}$  are the total masses of the volume-filling gas and the system of clouds in the CGM. For drag forces to be important, the volume-filling phase needs to significantly exceed the total cloud mass. The amount of mass in a volume-filling phase is uncertain in typical Milky Way-like galaxies \citep[e.g. ][]{miller13}, but appears to be significant in the halos of more massive galaxies \citep{PlanckCollaboration13,greco15}. 

We also find that the ratio $\Delta v/\sigma_{los}$ is typically large (mean value of $\sim$ 4, see Figure~11) interior to the virial radius. This is true in both the blue and red galaxies. This result is inconsistent with at least one simple and otherwise plausible model in which a single line-of-sight through the CGM intersects many clouds with distinct locations and a wide range of line-of-sight velocities (e.g. a sea of clouds orbiting in the halo potential). Instead it implies that a typical line-of-sight through the CGM is intersecting a coherent structure (a cloud, sheet, or filament). Current constraints on the size of these structures (which are based on models in which the gas is photoionized by the meta-galactic background) are rather weak, but characteristic path-lengths are of-order 1 to 10 kpc (Stocke et al. 2013; Werk et al. 2014).

Despite many of the similarities between the CGM in red and blue galaxies noted above, we do find significant differences. First, in terms of the radial distributions of the \Lya and \ion{Si}{3} equivalent widths, the red galaxies have lower normalizations for the exponential fits, reflecting a patchy distribution of absorbers. There is also a significant fraction of the blue galaxies showing large velocity differences between the radial velocity of the absorber and central galaxy ($\Delta v$ up to 500 \kms). Taken together with the COS-Halos results on the presence (absence) of highly ionized gas \citep[seen as \ion{O}{6} absorbers][]{tumlinson11b} in blue (red) galaxies, there are therefore real differences between the CGM surrounding star-forming and quiescent galaxies. Unfortunately, as discussed in Section 3.1, the data from the COS-GASS do not cover \ion{O}{6}, and thus we are not able to do the same analysis as done by \citet{tumlinson11b} for the combined sample.

The interpretation of these differences is not straightforward. The direction of the causal connection between the properties of the CGM and the galaxy could be in either direction (or both). The CGM properties could be influenced by feedback from the galaxy, and the amount and nature of this could be significantly different between the blue and red galaxies (e.g. feedback from massive stars and supernovae {\it vs.} feedback associated with AGN-driven radio sources (``radio mode'"). Alternatively, or in addition, the star-forming properties of the galaxies may be driven by the different observed properties of the CGM. 
 
One the main motivations of this paper were to understand whether and how the properties of the CGM in star-forming and quiescent galaxies relate to the cessation of star-formation in the latter. The results we have presented seem to imply rather subtle differences between the two types of galaxies. Here, we will argue that the apparently subtle differences could have significant implications.
 
Let us consider a simple model in which the star-formation rate in a given galaxy is proportional to the total mass of the system of CGM clouds traced by the absorption-lines ($M_{cs}$) divided by the timescale for these clouds to be transported from the CGM to the galaxy ($t_{in} \propto R_{vir}/v_{in}$).  The results in Figure 2 imply that $M_{cs} \propto f_c R_{vir}^2$, where $f_c$ is the fraction of the cross-sectional area of the CGM ($\pi R_{vir}^2$) covered by these clouds. Empirically, we see no dependence of the cloud kinematics on the mass of the dark matter halo. Hence, we will assume that $v_{in}$ is likewise independent of $M_{halo}$.   Taken together this implies that the specific star-formation rate scales as $sSFR \propto f_c R_{vir}^2/R_{vir} M_* \propto f_c R_{vir}/M_*$.  Using our adopted scaling between $M_*$ and $R_{vir}$, the predicted value of sSFR drops by a factor of ten with increasing stellar mass over the range in $M_* = 10^{10.0}$ to $10^{11.5}$ M$_{\odot}$, even if $f_c$ does not change. Given that $f_c$ is a factor of 3 smaller in the red galaxies (Figures 12 and 14), this simple model then predicts a difference in the sSFR between the lowest-mass blue galaxies and highest-mass red galaxies in our sample of a factor of 30.
 
Of course this model is purely phenomenological and does not explain the scaling of $M_c$ with $R_{vir}$ or the invariance in $v_{in}$. These relations are simply based on the observations. One possible interpretation would be that the mass of the population of clouds relative to that of the hot diffuse volume-filling CGM phase drops as a function of $M_{halo}$.  This reduces the normalized mass of the reservoir of clouds ($M_{cs}/M_*$) and increases the transport time of the clouds to the galaxy with increasing mass due to increasing drag forces.  Such an idea is at least qualitatively consistent with the simple paradigm of a transition from predominantly cold accretion to a quasi-hydrostatic hot CGM as halo and galaxy mass increases.

\section{Summary\label{sec:conclusion}} 

We have presented the analysis of a comprehensive data set combining the COS-GASS and COS-Halos samples to probe the CGM of low-z galaxies spanning a stellar mass range of almost two orders-of-magnitude centered on the characteristic mass ($\sim 10^{10.5} \rm M_{\odot}$) at which the galaxy population transitions from mostly blue, star-forming galaxies to red, quiescent ones. These two surveys cover similar ranges in stellar masses and dark halo virial radii ($\rm R_{vir}$). In addition, the combined sample uniformly samples a large range of radial distances from 0.02 to 1.3~$\rm R_{vir}$. The COS-GASS survey primarily samples the outer CGM and COS-Halos survey primarily samples the inner CGM. We characterized the CGM properties, including its radial profile, its kinematics, and its correlation with the global properties of the galaxies. In particular we have divided the sample into blue galaxies (with sSFR $\rm >~10^{-11}~ M_{\odot}yr^{-1}$) and red galaxies (with lower sSFR).   

In this analysis, we discussed the \Lya $\lambda$1215$\rm \AA$ and \ion{Si}{3} $\lambda$1206$\rm \AA$ transitions tracing intermediate ionization gas. \ion{Si}{3} was chosen as it is the strongest metal transition detected in the combined data set. The typical detection limit for the COS-GASS sample is $\sim \rm 50~m\AA$ corresponding to 3$\sigma$ uncertainty in the data. In the combined sample, the detection rates of \Lya and \ion{Si}{3} were 91\% and $\sim$50\% respectively. 

Based on the analysis of the combined sample we conclude the following:

\begin{itemize}

\item[1.] The radial distribution of the equivalent width of \Lya as a function of normalized impact parameter ($\rm \rho/R_{vir}$) can be expressed as an exponential. The scale-lengths are similar for the red and blue galaxies (0.75 and 0.72 $\rm R_{vir}$ respectively). The radial distribution of equivalent width of \ion{Si}{3} can also be expressed as an exponential with a scale-length of 0.36 (0.33) $\rm R_{vir}$ for the red (blue) galaxies. The detection rate of \ion{Si}{3} drops to almost zero beyond about 0.8 $\rm R_{vir}$.

\item[2.] The blue galaxies show a relatively uniform radial distribution of \Lya absorbers, implying an areal covering fraction of nearly 100\% in the CGM. In contrast, the \Lya absorbers have a much less uniform radial distribution in the CGM of the red galaxies, suggesting a patchy distribution with smaller areal covering fractions. These differences are reflected in the overall normalization of the radial distribution of equivalent widths, which is higher for the blue galaxies (by 0.45 dex). Similar results were found for \ion{Si}{3}, but are restricted to the region interior to 0.8 $\rm R_{vir}$ (where \ion{Si}{3} is detected).

\item[3.] We found a significant positive correlation between the equivalent width of \Lya and the star-formation rate (at the 99.8\% confidence level).  The correlation is even more significant for normalized quantities: the impact- parameter-corrected equivalent with of \Lya  ([Log W - $\rm \overline{Log~W}]_{Ly\alpha}$) and the specific SFR (SFR/M$_{\star}$) were found to correlate at the 99.99\% confidence level. Similar results were found for \ion{Si}{3}.

\item[4.] We found the velocity distribution of the centroids of the majority of the \Lya and \ion{Si}{3} to generally lie within $\sim$150~\kms of the systemic velocity of the galaxy. These velocities are smaller than the escape velocity, thus suggesting the gas seen in absorption is the gravitationally bound within the halo. The metal-line transitions are also found mostly within $\pm$ 100~\kms of \Lya absorbers, although not all strong ($\rm >0.3\AA$) \Lya absorbers showed associated \ion{Si}{3}.

\item[5.] We find that the velocity offset between the \Lya centroid and the systemic velocity ($\Delta v$) is usually significantly larger than the line-of-sight velocity dispersion of the \Lya line ($\sigma_{los}$). The mean ratio $\Delta v/\sigma_{los} \sim4$.

\item[6.] We find no dependence of the kinematic properties of the CGM ($\Delta v$ or $\sigma_{los}$) on the galaxy halo mass (virial velocity). This is surprising, as the sample spans ranges of about $10^{2}$ in halo mass and $\sim$5 in $v_{vir}$.

\item[7.] We found that the kinematic properties of the CGM are generally similar between the blue and red galaxies. However, while the majority of both the blue and red galaxies have $\Delta v <$ 100 \kms, the distribution of $\Delta v$ for the blue galaxies shows a pronounced tail out to values as high as 500 \kms. 

\item[8.] We found a significant change in the CGM kinematics at about a radius of 1.0 (0.7) $\rm R_{vir}$ for the blue (red) galaxies. In the outer CGM $\Delta v$ for the \Lya absorbers is always less than 150 \kms, while the distributions of $\Delta v$ show tails out to values as high as 500 \kms in the inner CGM. In addition, $\sigma_{los}$ is higher on average in the outer CGM for both the blue and red galaxies. These two results lead to a corresponding decrease in $\Delta v/\sigma_{los}$ in the outer CGM.

\end{itemize}

The combined COS-GASS and COS-Halos sample has allowed us to conduct a comprehensive study of the connection of the properties of the CGM with those of the stellar body of the galaxy. We think that three of the specific results from above are particularly noteworthy. First, the differences in the radial distributions of the \Lya {\it vs.} the \ion{Si}{3} absorbers, suggest that the inner CGM is being (or has at some time been) affected by feedback associated with massive stars and supernovae. This feedback has chemically-enriched the CGM. It is interesting that the kinematics of the inner and outer CGM also show differences, although it is not clear that these are related to feedback or to projection effects.

Secondly, the fact that the typical ratio of the velocity offset to the line-of-sight velocity dispersion for the \Lya absorption-lines is so large is an important clue as to the structure of the CGM. It implies that a line-of-sight through the CGM does not intersect a whole sea of many clouds orbiting in the halo, but is rather passing through a coherent structure (cloud, sheet, filament). Model-dependent estimates imply a path-length of order 1 to 10 kpc for these structures.

Thirdly, the independence of the kinematic properties of the warm CGM on the halo mass is quite remarkable. This implies that, even though the observed absorption-line systems are mostly gravitationally bound to the halo, simple gravitational forces alone do not adequately explain the CGM dynamics. For the massive red galaxies, the ``sub-virial"" velocities could be understood if the absorbers represent material cooling and condensing out of (or suffering drag as they move through) a hot volume-filling phase that is supported hydrostatically against gravity. 

A major motivation of this study was to understand how and why galaxies in this stellar mass regime exhibit the color bimodality stemming from a suppression/cessation of star-formation in some of them. Since the CGM is the interface through which galaxies could exchange gas and energy that is required to form stars (or is expelled as a result of star-formation), the CGM properties could hold clues as to how this process of gas delivery may be disrupted leaving some galaxies deprived of fuel to form stars. 
 We have explored a simple scenario in which the star-formation rate in a galaxy is proportional to the total mass of CGM clouds divided by an inflow time. We then show that the empirical results on the independence of CGM kinematic properties on halo mass and the smaller covering factor in the CGM in the red galaxies would imply a drop in the specific star formation rate by about a factor of 30 over the stellar mass range from 10$^{10}$ to 10$^{11.5}$ M$_{\odot}$.

In any event, we believe the data we have presented here provide a valuable observational resource for on-going and future numerical simulations that try to reproduce CGM properties such as \Lya and metal-line column density profiles, covering fraction, and dynamics, such as line-widths and velocity spreads \citep[for example][and references therein]{hummels12, stinson12, shen13, ford13, ford14, ford16, liang16, kauffmann16, fielding16}. The ultimate goal is understanding the role of the CGM in the evolution of galaxies.

\vspace{.5cm}
\acknowledgements 
We thank the referee for his useful comments.
We thank Hsiao-Wen Chen, Cameron Hummels, Colin Norman, Josh Peek, Molly Peeples, Jason X. Prochaska, John Stocke, and Jessica Werk for helpful discussions.
This work is based on observations with the NASA/ESA Hubble Space Telescope, which is operated by the Association of Universities for Research in Astronomy, Inc., under NASA contract NAS5-26555. SB and TH were supported by grant HST GO 12603. BC gratefully acknowledges support from the Australian Research Council's Future Fellowship (FT120100660) and Discovery Project (DP150101734) funding schemes."

This project also made use of SDSS data. Funding for the SDSS and SDSS-II has been provided by the Alfred P. Sloan Foundation, the Participating Institutions, the National Science Foundation, the U.S. Department of Energy, the National Aeronautics and Space Administration, the Japanese Monbukagakusho, the Max Planck Society, and the Higher Education Funding Council for England.  The SDSS Web Site is http://www.sdss.org/. 
The SDSS is managed by the Astrophysical Research Consortium for the Participating Institutions. The Participating Institutions are the American Museum of Natural History, Astrophysical Institute Potsdam, University of Basel, University of Cambridge, Case Western Reserve University, University of Chicago, Drexel University, Fermilab, the Institute for Advanced Study, the Japan Participation Group, Johns Hopkins University, the Joint Institute for Nuclear Astrophysics, the Kavli Institute for Particle Astrophysics and Cosmology, the Korean Scientist Group, the Chinese Academy of Sciences (LAMOST), Los Alamos National Laboratory, the Max-Planck-Institute for Astronomy (MPIA), the Max-Planck-Institute for Astrophysics (MPA), New Mexico State University, Ohio State University, University of Pittsburgh, University of Portsmouth, Princeton University, the United States Naval Observatory, and the University of Washington.

{\it Facilities:}  \facility{Sloan ()} \facility{HST (COS)}

\bibliographystyle{apj}	        
\bibliography{myref_bibtex}		

\clearpage

\clearpage
\begin{deluxetable}{ccccc ccccc   ccccc ccccc}  
\tabletypesize{\scriptsize}
\tablecaption{Description of Galaxy Properties for the COS-GASS survey$^a$. \label{tbl-galaxy}}
\tablewidth{0pt}
\tablehead{
\colhead{Galaxy} &\colhead{GASS ID} & \colhead{RA} & \colhead{Dec} &\colhead{$\rm z_{gal}$} & \colhead{$\rm M_{\star} $} & \colhead{$\rm M_{halo}^b$} & \colhead{$\rm R_{vir}^c$} & \colhead{sSFR}   & \colhead{Color$^d$}  & \colhead{$\rm v_{esc,Rvir}^e$} \\
\colhead{}   &\colhead{}   & \colhead{}  & \colhead{}  & \colhead{} &\colhead{($\rm Log~M_{\odot}$)} &\colhead{($\rm Log~M_{\odot}$)} & \colhead{(kpc)}  &\colhead{($\rm Log~yr^{-1}$)}  & \colhead{} & \colhead{($\rm km~s^{-1}$)} }
\startdata
J0159+1346  &          3936  &  29.941 & 13.781 & 0.0441 & 10.1  &  11.4  &  153  &  -9.5  &  Blue & 114  \\
J0808+0512  &         19852  &  122.068 & 5.216 & 0.0308 & 10.8  &  12.2  &  296  &  -12.0  &  Red & 217  \\
J0852+0309  &          8096  &  133.229 & 3.152 & 0.0345 & 10.3  &  11.5  &  166  &  -10.1  &  Blue & 122  \\
J0908+3234  &         22391  &  137.232 & 32.576 & 0.0490 & 10.5  &  11.9  &  232  &  -12.3  &  Red & 174  \\
J0914+0836  &         20042  &  138.684 & 8.601 & 0.0468 & 10.0  &  11.3  &  147  &  -9.6  &  Blue & 110  \\
J0930+2853  &         32907  &  142.538 & 28.898 & 0.0349 & 10.5  &  11.6  &  184  &  -10.7  &  Blue & 136  \\
J0931+2632  &         53269  &  142.817 & 26.550 & 0.0458 & 11.0  &  12.4  &  345  &  -12.6  &  Red & 258  \\
J0936+3204  &         33214  &  144.101 & 32.079 & 0.0269 & 10.3  &  11.8  &  217  &  -11.7  &  Red & 158  \\
J0937+1658  &         55745  &  144.292 & 16.977 & 0.0278 & 10.9  &  12.0  &  263  &  -10.3  &  Blue & 192  \\
J0951+3537  &         22822  &  147.937 & 35.622 & 0.0270 & 10.6  &  11.7  &  197  &  -10.4  &  Blue & 143  \\
J0958+3204  &         33737  &  149.714 & 32.073 & 0.0270 & 10.7  &  12.1  &  272  &  -12.7  &  Red & 198  \\
J1002+3238  &         33777  &  150.711 & 32.645 & 0.0477 & 10.1  &  11.7  &  191  &  -11.9  &  Red & 143  \\
J1013+0501  &          8634  &  153.352 & 5.025 & 0.0464 & 10.1  &  11.4  &  153  &  -10.8  &  Blue & 115  \\
J1032+2112  &         55541  &  158.196 & 21.216 & 0.0429 & 10.6  &  11.7  &  202  &  -10.1  &  Blue & 150  \\
J1051+1245  &         23419  &  162.827 & 12.757 & 0.0400 & 10.4  &  11.5  &  175  &  -10.0  &  Blue & 130  \\
J1059+0517  &          9109  &  164.811 & 5.292 & 0.0353 & 11.1  &  12.6  &  387  &  -11.9  &  Red & 284  \\
J1100+1210  &         23457  &  165.048 & 12.171 & 0.0354 & 10.1  &  11.4  &  154  &  -10.7  &  Blue & 114  \\
J1100+1043  &         23477  &  165.200 & 10.728 & 0.0360 & 11.1  &  12.3  &  313  &  -11.0  &  Blue & 231  \\
J1115+0241  &          5701  &  168.789 & 2.699 & 0.0442 & 10.7  &  11.8  &  218  &  -10.9  &  Blue & 162  \\
J1120+0410  &         12452  &  170.026 & 4.177 & 0.0492 & 10.8  &  12.2  &  296  &  -12.1  &  Red & 222  \\
J1122+0314  &          5872  &  170.642 & 3.244 & 0.0446 & 10.5  &  11.9  &  239  &  -12.0  &  Red & 178  \\
J1127+2657  &         48604  &  171.943 & 26.960 & 0.0334 & 10.6  &  11.7  &  201  &  -11.0  &  Blue & 147  \\
J1131+1553  &         29898  &  172.954 & 15.897 & 0.0364 & 10.2  &  11.7  &  199  &  -12.0  &  Red & 147  \\
J1132+1329  &         29871  &  173.052 & 13.492 & 0.0342 & 10.2  &  11.4  &  158  &  -9.7  &  Blue & 116  \\
J1142+3013  &         48994  &  175.575 & 30.230 & 0.0322 & 10.7  &  11.8  &  222  &  -10.4  &  Blue & 163  \\
J1155+2921  &         49433  &  178.903 & 29.351 & 0.0458 & 10.5  &  11.6  &  180  &  -10.3  &  Blue & 135  \\
J1241+2847  &         50550  &  190.367 & 28.791 & 0.0350 & 10.3  &  11.5  &  166  &  -10.0  &  Blue & 123  \\
J1251+0551  &         13074  &  192.894 & 5.864 & 0.0486 & 10.9  &  12.0  &  242  &  -10.4  &  Blue & 182  \\
J1305+0359  &         13159  &  196.356 & 3.992 & 0.0437 & 10.4  &  11.5  &  172  &  -10.8  &  Blue & 128  \\
J1315+1525  &         26936  &  198.855 & 15.423 & 0.0266 & 10.7  &  12.1  &  283  &  -12.3  &  Red & 206  \\
J1317+2629  &         51025  &  199.440 & 26.486 & 0.0450 & 10.3  &  11.4  &  162  &  -10.4  &  Blue & 121  \\
J1325+2714  &         51161  &  201.345 & 27.249 & 0.0345 & 10.1  &  11.4  &  156  &  -9.8  &  Blue & 115  \\
J1348+2453  &         38018  &  207.142 & 24.891 & 0.0297 & 10.1  &  11.3  &  153  &  -10.5  &  Blue & 112  \\
J1354+2433  &         44856  &  208.546 & 24.556 & 0.0286 & 10.1  &  11.6  &  191  &  -11.8  &  Red & 139  \\
J1404+3357  &         31172  &  211.122 & 33.953 & 0.0264 & 10.3  &  11.8  &  211  &  -12.3  &  Red & 154  \\
J1406+0154  &          7121  &  211.678 & 1.915 & 0.0472 & 10.2  &  11.7  &  202  &  -11.8  &  Red & 152  \\
J1427+2629  &         45940  &  216.954 & 26.484 & 0.0325 & 10.4  &  11.9  &  225  &  -12.0  &  Red & 165  \\
J1430+0323  &          9615  &  217.508 & 3.398 & 0.0333 & 10.2  &  11.7  &  197  &  -11.1  &  Red & 145  \\
J1431+2440  &         38198  &  217.894 & 24.682 & 0.0378 & 10.7  &  12.1  &  261  &  -12.7  &  Red & 193  \\
J1454+3050  &         42191  &  223.516 & 30.846 & 0.0320 & 10.1  &  11.4  &  155  &  -9.8  &  Blue & 114  \\
J1502+0649  &         41743  &  225.517 & 6.823 & 0.0462 & 10.5  &  11.6  &  180  &  -10.2  &  Blue & 135  \\
J1509+0704  &         41869  &  227.340 & 7.078 & 0.0414 & 10.1  &  11.4  &  155  &  -9.6  &  Blue & 115  \\
J1515+0701  &         42025  &  228.781 & 7.021 & 0.0367 & 10.9  &  12.3  &  314  &  -11.9  &  Red & 231  \\
J1541+2813  &         28365  &  235.344 & 28.230 & 0.0321 & 10.4  &  11.5  &  173  &  -9.6  &  Blue & 127  \\
J1544+2740  &         28317  &  236.034 & 27.673 & 0.0316 & 10.1  &  11.6  &  191  &  -12.1  &  Red & 140  \\
\enddata
\tablenotetext{a}{Details on the COS-Halos survey can be found in the published work by \citet{tumlinson13, werk13}.}
\tablenotetext{b}{Using prescription from \citet{kravtsov14}.}
\tablenotetext{c}{Using prescription from \citet{liang14}.}
\tablenotetext{d}{Galaxies with sSFR $>\rm 10^{-11}~yr^{-1}$ are defined as blue galaxies.  Galaxies with sSFR below this value are defined as red galaxies.}
\tablenotetext{e}{Escape velocity at the virial radii probed by the QSO sightline assuming a NFW profile for the galaxy's dark matter distribution.}
\end{deluxetable}

\clearpage
\LongTables 
\begin{landscape}
\begin{deluxetable}{ccccc ccccc   ccccc ccccc}  
\tabletypesize{\scriptsize}
\tablecaption{Description of QSO Sightlines and Absorption Line Measurements for the COS-GASS survey. \label{tbl-sightlines}}
\tablewidth{0pt}
\tablehead{
\colhead{QSO} & \colhead{RA$_{\rm QSO}$} & \colhead{Dec$_{\rm QSO}$} & \colhead{z$_{\rm QSO}$} & \colhead{$\rho$} & \colhead{$\rm \rho/R_{vir}$} & \colhead{$\rm \Theta^a$}   & \colhead{$\rm W_{Ly\alpha}^b$} &  \colhead{$\rm \Delta V_{Ly\alpha}^c$}  &\colhead{$v_{\rm Ly\alpha}^d$} &  \colhead{$b_{\rm Ly\alpha}^d$} &  \colhead{$\rm W_{SiIII1206}$}   & \colhead{$\rm \Delta V_{SiIII1205}^c$}  & \colhead{$v_{\rm SiIII}^d$}  & \colhead{$b_{\rm SiIII}^d$} \\
\colhead{}    &   \colhead{}             & \colhead{}                & \colhead{}              & \colhead{(kpc)}  & \colhead{}                   & \colhead{}                 & \colhead{($\rm \AA$)}        & \colhead{($\rm km~s^{-1}$)}  & \colhead{($\rm km~s^{-1}$)}  & \colhead{($\rm km~s^{-1}$)} & \colhead{($\rm \AA$)}    & \colhead{($\rm km~s^{-1}$)}  & \colhead{($\rm km~s^{-1}$)}}
\startdata
J0159+1345  &  29.971  &  13.765  &  0.504  &  102  &  0.7  &  64 & 1.501$\pm$0.023 &  -150 $-$ 380
 & 93,358  & 85,14 &  $-$ &  $-$  & $-$  &$-$\\
J0808+0514  &  122.162  &  5.244  &  0.361  &  215  &  0.7  &  7 &  $-$ &  $-$  & $-$  & $-$ &  $<$0.050 &  $-$  & $-$  &$-$\\
J0852+0313  &  133.247  &  3.222  &  0.297  &  178  &  1.1  &  67 & 0.113$\pm$0.015 &  0 $-$ 110
 & 50  & 49 &  $<$0.051 &  $-$  & $-$  &$-$\\
J0909+3236  &  137.276  &  32.608  &  0.809  &  170  &  0.7  &  21 & 0.090$\pm$0.014 &  50 $-$ 200
 & 101  & 71 &  $<$0.041 &  $-$  & $-$  &$-$\\
J0914+0837  &  138.632  &  8.629  &  0.649  &  189  &  1.3  &  69 & 0.104$\pm$0.020 &  -50 $-$ 80
 & 30  & 45 &  $<$0.065 &  $-$  & $-$  &$-$\\
J0930+2848  &  142.508  &  28.816  &  0.487  &  214  &  1.2  &  42 &  $<$0.126 &  $-$  & $-$  & $-$ &  $<$0.108 &  $-$  & $-$  &$-$\\
J0931+2628  &  142.820  &  26.480  &  0.778  &  226  &  0.7  &  77 & 0.114$\pm$0.013 &  150 $-$ 250
 & 200  & 55 &  $<$0.043 &  $-$  & $-$  &$-$\\
J0936+3207  &  144.016  &  32.119  &  1.150  &  160  &  0.7  &  0$^e$ &  $<$0.113 &  $-$  & $-$  & $-$ &  $<$0.086 &  $-$  & $-$  &$-$\\
J0937+1700  &  144.279  &  17.006  &  0.506  &  63  &  0.2  &  64 & 0.135$\pm$0.021 &  -300 $-$ -150
 & -204  & 67 &  $<$0.056 &  $-$  & $-$  &$-$\\
J0951+3542  &  147.850  &  35.714  &  0.398  &  226  &  1.1  &  3 & 0.839$\pm$0.014 &  -90 $-$ 200
 & 64  & 59 &  $-$ &  $-$  & $-$  &$-$\\
J0959+3203  &  149.812  &  32.066  &  0.564  &  162  &  0.6  &  0$^e$ & 0.420$\pm$0.016 &  -320 $-$ -100
 & -238,-144  & 33,19 & 0.060$\pm$0.013 & 
-300 $-$ -200
 &-256 &12\\
J1002+3240  &  150.727  &  32.678  &  0.829  &  119  &  0.6  &  40 &  $<$0.063 &  $-$  & $-$  & $-$ &  $<$0.062 &  $-$  & $-$  &$-$\\
J1013+0500  &  153.325  &  5.009  &  0.266  &  102  &  0.7  &  0$^e$ & 0.445$\pm$0.024 &  -180 $-$ 50
 & -72  & 49 & 0.088$\pm$0.020 & 
-120 $-$ 0
 &-70 &20\\
J1033+2112  &  158.270  &  21.204  &  0.315  &  214  &  1.1  &  45 & 0.437$\pm$0.033 &  -100 $-$ 95
 & -1  & 50 &  $<$0.065 &  $-$  & $-$  &$-$\\
J1051+1247  &  162.857  &  12.796  &  1.281  &  140  &  0.8  &  73 & 0.781$\pm$0.022 &  -180 $-$ 140
 & -37  & 75 &  $-$ &  $-$  & $-$  &$-$\\
J1059+0519  &  164.795  &  5.327  &  0.754  &  95  &  0.2  &  30 & 0.270$\pm$0.022 &  -100 $-$ 100
 & 6  & 44 &  $<$0.062 &  $-$  & $-$  &$-$\\
J1059+1211  &  164.984  &  12.198  &  0.993  &  171  &  1.1  &  61 & 0.225$\pm$0.014 &  -150 $-$ 0
 & -61  & 26 &  $<$0.045 &  $-$  & $-$  &$-$\\
J1100+1046  &  165.199  &  10.770  &  0.422  &  108  &  0.3  &  0$^e$ &  $-$ &  $-$  & $-$  & $-$ &  $-$ &  $-$  & $-$  &$-$\\
J1115+0237  &  168.782  &  2.633  &  0.567  &  209  &  1.0  &  83 & 0.195$\pm$0.020 &  -100 $-$ 100
 & 2  & 37 & 0.065$\pm$0.012 & 
-20 $-$ 30
 &4 &20\\
J1120+0413  &  170.021  &  4.223  &  0.545  &  162  &  0.5  &  78 & 0.830$\pm$0.018 &  50 $-$ 390
 & 201,370  & 73,21 & 0.172$\pm$0.015 & 
150 $-$ 290
 &194 &26\\
J1122+0318  &  170.601  &  3.301  &  0.475  &  221  &  0.9  &  0$^e$ & 0.110$\pm$0.018 &  0 $-$ 150
 & 67  & 58 &  $-$ &  $-$  & $-$  &$-$\\
J1127+2654  &  171.902  &  26.914  &  0.379  &  140  &  0.7  &  26 & 0.705$\pm$0.021 &  -250 $-$ 100
 & -28,-178  & 54,30 & 0.051$\pm$0.017 & 
-80 $-$ 20
 &-34 &12\\
J1131+1556  &  172.905  &  15.946  &  0.183  &  176  &  0.9  &  0$^e$ &  $-$ &  $-$  & $-$  & $-$ &  $-$ &  $-$  & $-$  &$-$\\
J1132+1335  &  173.044  &  13.586  &  0.201  &  230  &  1.5  &  5 & 0.319$\pm$0.017 &  -40 $-$ 130
 & 57  & 55 &  $<$0.043 &  $-$  & $-$  &$-$\\
J1142+3016  &  175.551  &  30.270  &  0.481  &  104  &  0.5  &  50 & 0.886$\pm$0.023 &  -200 $-$ 170
 & 3  & 67 & 0.227$\pm$0.019 & 
-150 $-$ 50
 &-33 &53\\
J1155+2922  &  178.970  &  29.377  &  0.520  &  208  &  1.2  &  1 & 0.742$\pm$0.023 &  -200 $-$ 230
 & 72,-150  & 60,13 & 0.056$\pm$0.015 & 
40 $-$ 160
 &92 &66\\
J1241+2852  &  190.374  &  28.870  &  0.589  &  198  &  1.2  &  40 & 0.211$\pm$0.020 &  -120 $-$ 170
 & 33  & 95 &  $<$0.040 &  $-$  & $-$  &$-$\\
J1251+0554  &  192.853  &  5.906  &  1.377  &  200  &  0.8  &  57 & 0.409$\pm$0.021 &  -20 $-$ 180
 & 80  & 51 &  $<$0.063 &  $-$  & $-$  &$-$\\
J1305+0357  &  196.351  &  3.959  &  0.545  &  103  &  0.6  &  11 & 0.821$\pm$0.016 &  -160 $-$ 180
 & 67,-43  & 48,59 &  $-$ &  $-$  & $-$  &$-$\\
J1315+1525  &  198.938  &  15.432  &  0.448  &  155  &  0.5  &  17 & 0.405$\pm$0.018 &  -50 $-$ 170
 & 64  & 48 & 0.131$\pm$0.017 & 
0 $-$ 150
 &49 &48\\
J1318+2628  &  199.508  &  26.475  &  1.234  &  198  &  1.2  &  86 & 0.184$\pm$0.032 &  -120 $-$ 120
 & 0  & 50 &  $-$ &  $-$  & $-$  &$-$\\
J1325+2717  &  201.266  &  27.289  &  0.522  &  199  &  1.3  &  55 &  $-$ &  $-$  & $-$  & $-$ &  $<$0.095 &  $-$  & $-$  &$-$\\
J1348+2456  &  207.093  &  24.947  &  0.293  &  153  &  1.0  &  81 & 0.474$\pm$0.035 &  -230 $-$ 0
 & -97  & 76 &  $<$0.073 &  $-$  & $-$  &$-$\\
J1354+2430  &  208.604  &  24.502  &  1.878  &  155  &  0.8  &  78 & 0.545$\pm$0.034 &  -170 $-$ 50
 & -97,-5  & 37,26 &  $<$0.084 &  $-$  & $-$  &$-$\\
J1404+3353  &  211.118  &  33.895  &  0.549  &  111  &  0.5  &  57 & 0.749$\pm$0.027 &  -150 $-$ 150
 & -26  & 75 & 0.177$\pm$0.024 & 
0 $-$ 150
 &37 &45\\
J1406+0157  &  211.732  &  1.954  &  0.427  &  222  &  1.1  &  67 &  $-$ &  $-$  & $-$  & $-$ &  $<$0.061 &  $-$  & $-$  &$-$\\
J1427+2632  &  216.898  &  26.537  &  0.364  &  170  &  0.8  &  0$^e$ &  $-$ &  $-$  & $-$  & $-$ &  $<$0.078 &  $-$  & $-$  &$-$\\
J1429+0321  &  217.420  &  3.357  &  0.253  &  231  &  1.2  &  0$^e$ & 0.807$\pm$0.027 &  -150 $-$ 250
 & -39,109  & 68,104 & 0.052$\pm$0.017 & 
-50 $-$ 50
 &-21 &42\\
J1431+2442  &  217.858  &  24.706  &  0.407  &  110  &  0.4  &  18 & 0.569$\pm$0.015 &  0 $-$ 220
 & 73,156  & 40,28 &  $<$0.048 &  $-$  & $-$  &$-$\\
J1454+3046  &  223.601  &  30.783  &  0.465  &  223  &  1.4  &  37 & 0.472$\pm$0.035 &  -50 $-$ 160
 & 57  & 47 &  $<$0.079 &  $-$  & $-$  &$-$\\
J1502+0645  &  225.517  &  6.754  &  0.288  &  224  &  1.2  &  80 & 0.438$\pm$0.013 &  -150 $-$ 110
 & 12,-55  & 38,72 &  $<$0.032 &  $-$  & $-$  &$-$\\
J1509+0702  &  227.368  &  7.043  &  0.418  &  130  &  0.8  &  62 & 0.956$\pm$0.022 &  -275 $-$ 130
 & 49,-18,-215  & 29,84,30 & 0.137$\pm$0.011 & 
-275 $-$ -195
 &-239 &21\\
J1515+0657  &  228.781  &  6.952  &  0.268  &  180  &  0.6  &  14 & 0.270$\pm$0.023 &  -500 $-$ -300
 & -367  & 43 &  $<$0.061 &  $-$  & $-$  &$-$\\
J1541+2817  &  235.340  &  28.285  &  0.376  &  128  &  0.7  &  0$^e$ & 0.864$\pm$0.011 &  -520 $-$ -250
 & -363  & 90 &  $<$0.039 &  $-$  & $-$  &$-$\\
J1544+2743  &  236.114  &  27.723  &  0.163  &  196  &  1.0  &  55 & 0.191$\pm$0.022 &  50 $-$ 210
 & 126  & 69 &  $<$0.064 &  $-$ & $-$  &$-$
\enddata
\tablenotetext{a}{Orientation of the QSO sightlines with respect to the disk of the galaxies. The values are based on SDSS r-band photometric measurements.}
\tablenotetext{b}{Limiting equivalent width denotes 3$\sigma$ uncertainity.}
\tablenotetext{c}{Full width of the absorption feature in the rest-frame of the galaxy.}
\tablenotetext{d}{Centroid and b-value of the multiple components of the \Lya and Si~III absorption feature as estimated via Voigt profile fit. These are printed in the order of the strength of the component. Redshift of the absorber $\rm z_{abs} = z_{gal} + { \it v_{transition}}/c$, where  $\rm z_{gal}$ is the systemic redshift of the galaxy (from Table~1) and c is the speed of light in vacuum. }
\tablenotetext{e}{Face-on galaxies.}
\end{deluxetable}
\clearpage
\end{landscape}

\end{document}